\documentclass[superscriptaddress,groupedaddress,nofootnoteinbib,12pt]{article}  

\usepackage{graphicx}  
\usepackage{dcolumn}   
\usepackage{bm}        
\usepackage{jcappub}

\def\be{\begin{equation}}
\def\ee{\end{equation}}
\def\ba{\begin{eqnarray}}
\def\ea{\end{eqnarray}}
\def\beq{\begin{eqnarray}}
\def\eeq{\end{eqnarray}}

\def\mpl{M_{\rm Pl}}

\def\d{\mathrm{d}}

\def\L*{{\cal L}_*}
\def\L{\mathcal{L}}
\def\({\left(}
\def\){\right)}

\def\nn{\nonumber}

\def\stu{St\"uckelberg }

\def\<{\langle}
\def\>{\rangle}

\def\cs2{c_{s}^{2}}

 \def\ep{\varepsilon}

 \def\om{\omega}
 \def\Om{\Omega}

 \def\wed{\wedge}

 \def\be   {\begin{equation}}   \def\ee   {\end{equation}}

 \def\ba  {\begin{eqnarray}}   \def\ea  {\end{eqnarray}}

\begin{document}

\title{Cutoff for Extensions of Massive Gravity and Bi-Gravity}
\author{Andrew Matas}
\affiliation{CERCA/Department of Physics, Case Western Reserve University, 10900 Euclid Ave, Cleveland, OH 44106, USA}
\emailAdd{andrew.matas@case.edu}

\abstract{Recently there has been interest in extending ghost-free massive gravity, bi-gravity, and multi-gravity by including non-standard kinetic terms and matter couplings. We first review recent proposals for this class of extensions, emphasizing how modifications of the kinetic and potential structure of the graviton and modifications of the coupling to matter are related. We then generalize existing no-go arguments in the metric language to the vielbein language in second-order form. We give an ADM argument to show that the most promising extensions to the kinetic term and matter coupling contain a Boulware-Deser ghost. However, as recently emphasized, we may still be able to view these extensions as effective field theories below some cutoff scale. To address this possibility, we show that there is a decoupling limit where a ghost appears for a wide class of matter couplings and kinetic terms. In particular, we show that there is a decoupling limit where the linear effective vielbein matter coupling contains a ghost. Using the insight we gain from this decoupling limit analysis, we place an upper bound on the cutoff for the linear effective vielbein coupling. This result can be generalized to new kinetic interactions in the vielbein language in second-order form. Combined with recent results, this provides a strong uniqueness argument on the form of ghost-free massive gravity, bi-gravity, and multi-gravity.}

\maketitle

\section{Introduction}
In the past few years there has been a revival of interest in theories of Lorentz-invariant interacting massive spin-2 fields, for reviews see \cite{Hinterbichler:2011tt,deRham:2014zqa}. This class of theories includes ghost-free massive gravity \cite{deRham:2010ik,deRham:2010kj}, bi-gravity \cite{Hassan:2011zd}, and multi-gravity \cite{Hinterbichler:2012cn,Hassan:2012wt}. The theories generalize the Fierz-Pauli mass term \cite{Fierz:1939ix} to the non-linear level in a way that avoids the Boulware-Deser (BD) ghost \cite{Boulware:1973my}, for proofs see \cite{Hassan:2012qv, Mirbabayi:2011aa,Hassan:2011hr,Hassan:2011ea,Hassan:2011tf,Hassan:2011vm, deRham:2011qq,deRham:2011rn}, 
see also \cite{Kugo:2014hja} for a perspective coming from BRST. 
These theories are of interest for cosmology 
\cite{deRham:2010tw,
D'Amico:2011jj,vonStrauss:2011mq,Volkov:2011an,Comelli:2011zm,
Chiang:2012vh,Akrami:2012vf,Tasinato:2012ze,
DeFelice:2013bxa,Akrami:2013pna,Volkov:2013roa,Andrews:2013uca,Do:2013tea,
Bamba:2013hza,Konnig:2013gxa,Guarato:2013gba,
Konnig:2014dna,Goon:2014ywa,Comelli:2014bqa,Solomon:2014dua,DeFelice:2014nja, Yilmaz:2014lfa,Konnig:2014xva,Cusin:2014psa,Lagos:2014lca,deRham:2014gla,
Nersisyan:2015oha,Amendola:2015tua,Mazuet:2015pea,Johnson:2015tfa,
Cusin:2015pya,Akrami:2015qga,Fasiello:2015csa}  
(as well as the mass varying \cite{Huang:2012pe,Leon:2013qh,Cai:2012ag,Hinterbichler:2013dv,Wu:2013ii,Bamba:2013aca}, $F(R)$ \cite{Cai:2013lqa,Cai:2014upa}, quasi-dilaton \cite{D'Amico:2012zv,Gannouji:2013rwa,Gumrukcuoglu:2013nza}, and extended quasi-dilaton \cite{DeFelice:2013tsa,Mukohyama:2014rca,DeFelice:2013dua,Mukohyama:2013raa,
Kahniashvili:2014wua,Heisenberg:2015voa} extensions)
and have been applied to studying condensed matter systems using AdS/CFT \cite{Vegh:2013sk,Blake:2013owa,Blake:2013bqa,Davison:2013jba,Zeng:2014uoa}. It is also an interesting fundamental question to ask how spin-2 fields can interact consistently.
\\

It is often useful to consider these theories in the decoupling limit, using the \stu formalism introduced in massive gravity by \cite{Siegel:1993sk,ArkaniHamed:2002sp}. The decoupling limit serves multiple functions. First, it simplifies the theory by focusing on the interactions that arise at the lowest scale, which for ghost-free massive gravity occur at $\Lambda_3 \sim (m^2 \mpl)^{1/3}$. Second, the decoupling limit can diagnose the presence or absence of a BD ghost. If the BD ghost is present it typically manifests itself as higher derivatives on the \stu fields \cite{Creminelli:2005qk,Deffayet:2005ys}, giving rise to an Ostragradski instability \cite{Ostrogradski}. As shown in \cite{deRham:2010ik}, for ghost-free massive gravity after diagonalizing\footnote{There is one interaction in massive gravity that cannot be diagonalized with a local field redefinition, however the equations of motion are still second order.}, the scalar modes are described by Galileon interactions \cite{Nicolis:2008in}, 
which have second order equation of motion, reflecting the absence of the BD ghost in the full theory. Finally, the decoupling limit simplifies the discussion of the Vainshtein mechanism \cite{Vainshtein:1972sx}, which is needed in order to avoid the van Dam-Veltman-Zakharov (vDVZ) discontinuity \cite{vanDam:1970vg,Zakharov:1970cc}. For a review of the Vainshtein mechanism see \cite{Babichev:2013usa}. The Vainshtein mechanism is a strong coupling effect that occurs when the curvature $R\sim m^2$, or in the decoupling limit when the Galileon operators at $\Lambda_3$ become large. For a proposal of a description of the strong coupling dynamics in Galileon theories at energies above $\Lambda_3$, see \cite{Keltner:2015xda}. The complete decouping limit of ghost-free massive gravity including the vectors was derived in \cite{Ondo:2013wka,Gabadadze:2013ria}, and the decoupling limit for multi-gravity was discussed in \cite{Noller:2015eda}.
\\

In this work we shall focus on bi-gravity (and the limiting case of massive gravity), which can be described in terms of two vielbeins $e^a_\mu$ and $f^a_\mu$. A natural question is whether ghost-free bi-gravity is complete, in the sense that it is the most general theory describing one massive and one massless spin-2 field that avoids the BD ghost and exhibits a Vainshtein mechanism. It is known that the potential interactions cannot be generalized further without introducing a BD ghost \cite{deRham:2010ik}. However, given that the potential term breaks diffeomorphism invariance, it is not {\it a priori} obvious that bi-gravity needs to maintain the same the kinetic structure and matter couplings from General Relativity (GR). 

Indeed both of these possiblities have been considered in the literature. Kinetic interactions that are not of the Einstein-Hilbert form were studied in \cite{Hinterbichler:2013eza,Noller:2014ioa,Kimura:2013ika,deRham:2015rxa}, and matter couplings where the matter fields are coupled simultaneously to both vielbeins were considered in \cite{Hassan:2011zd,Hassan:2012wr,Akrami:2013ffa,Yamashita:2014fga,deRham:2014naa,
Hassan:2014gta,deRham:2014fha,Noller:2014sta,Heisenberg:2014rka,Hinterbichler:2015yaa, deRham:2015cha,Akrami:2014lja,Soloviev:2014eea}. Such extensions may admit interesting cosmological solutions \cite{Comelli:2015pua,Gumrukcuoglu:2015nua,Gumrukcuoglu:2014xba,
Solomon:2014iwa,Enander:2014xga,Gao:2014xaa}, and may be relevant for partially massless gravity \cite{Deser:1983mm,Deser:1983tm,Deser:2001pe,Deser:2001us,Deser:2001wx,
Deser:2001xr,Deser:2003gw,Deser:2004ji,Deser:2006zx,Deser:2013bs,Deser:2013uy,
deRham:2012kf,deRham:2013wv,Hassan:2012gz}, dark matter \cite{Blanchet:2015sra}, and for electrically charged spin-2 fields \cite{deRham:2014tga} (which could be useful for holographic applications \cite{Hartnoll:2008kx}). Furthermore, if such extensions that do not excite a BD ghost exist, then we should consider them as part of ghost-free bi-gravity.
\\

However, there are also several arguments that these extensions contain ghosts. In the metric language in four dimensions, the possibility of new ghost-free kinetic interactions in massive gravity was ruled out in \cite{deRham:2013tfa}. Since the metric and vielbein formulations need not give the same theory when considering modifications of the kinetic structure or matter couplings \cite{Hinterbichler:2013eza,deRham:2015rxa,deRham:2015cha}, it is conceivable that there can be new kinetic interactions in the vielbein language. In \cite{deRham:2015rxa}, it was shown that in the vielbein language in first-order form that ghosts arise. In this work we extend these no-go arguments to the vielbein language in second-order form. 

Recently, there have also been arguments showing that modifications of the matter coupling give rise to ghosts \cite{deRham:2014naa,Huang:2015yga,Heisenberg:2015iqa}. In fact, we will argue that these no-go statements are closely related. As we will discuss below, the uniqueness of the potential term itself places strong constraints on the matter coupling, and the no-go arguments for the kinetic terms also imply no-go statements for non-minimal matter couplings. 
\\

Collectively, these arguments (including the arguments given in this paper against kinetic terms in the vielbein formulation in second-order form) show that there are no strictly ghost-free modifications of the kinetic term and matter coupling. However it may still be possible to view these couplings in an effective field theory sense below some cutoff. In particular, as discussed in \cite{deRham:2014wfa}, if there are ghost-free genuine interactions coming from the matter coupling or kinetic interaction, it may be possible for these interactions to become strongly coupled while still remaining within the regime of validity of the effective theory. This requires that these interactions come in at a scale $\Lambda_{\rm s.c.}$ that we can consistently take to be below the cutoff of the effective field theory (defined as the scale at which new physics enters), in other words so long as we can take $\Lambda_{\rm s.c.}<\Lambda_{\rm c.o.}$.
This logic was applied to the linear effective vielbein matter coupling proposed in \cite{deRham:2014naa} (see also related discussions for the matter coupling in \cite{Huang:2015yga,Heisenberg:2015iqa} and for new kinetic interactions in \cite{Noller:2014ioa}), in which matter is minimally coupled to the vielbein
\be
v^a_\mu = \alpha e^a_\mu + \beta f^a_\mu.
\ee
In \cite{deRham:2014naa}, it was shown that this coupling is ghost-free perturbatively around the mini-superspace, and it was shown that there was a decoupling limit in which the theory was ghost-free. Thus treated as effective field theories, these models could potentially be of interest for cosmology. Following this logic, it is important not just to determine whether or not there is a ghost, but at what scale the ghost arises. In this work, we will consider a different decoupling limit which contains ghostly operators, and use the result of this analysis to place an upper bound on the cutoff of this matter coupling.
\\

\noindent
{\it Summary of Main Results.}
\begin{itemize}
\item By performing an Arnowitt-Deser-Misner (ADM) analysis \cite{Arnowitt:1962hi} in the vielbein language, we show that the most promising kinetic and matter coupling extensions lead to a BD ghost. For the matter coupling, our argument gives a different perspective on how the constraint structure is modified than the analysis of \cite{deRham:2015cha}.
\item For a generic matter coupling, there is a ghost already at the scale $\Lambda_3$. This is consistent with the recent results of \cite{Huang:2015yga,Heisenberg:2015iqa}, though we will use slightly different assumptions. 
\item For the linear effective vielbein coupling, we find there is a decoupling limit, different from the one considered in \cite{deRham:2014naa}, in which the following ghostly dimension nine operator appears
\be
\mathcal{O}_{(\partial B) (\partial \partial \pi) T} = \frac{\alpha \beta m}{\Lambda_3^6}F_{\mu \alpha} \partial^\alpha \partial_{\nu} \pi T_{(\eta)}^{\mu\nu},
\ee
where $T^{\mu\nu}_{(\eta)}$ is the matter stress energy tensor on a Minkowski background and $F_{\mu\nu} \equiv \partial_\mu B_\nu - \partial_\nu B_\mu$, and where $B_\mu$ and $\pi$ are the vector and scalar \stu fields. This operator appears in both the metric and vielbein formulations. It is possible to take a different scaling limit because the interactions from the matter coupling vanish in the decoupling limit considered in \cite{deRham:2014naa}, as we will show using a Galileon duality transformation \cite{deRham:2013hsa,deRham:2014lqa}.
\item Beyond the decoupling limit, the existence of this operator allows us to put an upper bound on the cutoff
\be
\Lambda_{\rm c.o.} \lesssim (\alpha \beta)^{-1/5} \left(m^3 \mpl^2 \right)^{1/5}, \ {\rm for\ the\ linear\ effective\ vielbein}.
\ee
It is likely this bound can be improved by considering higher dimension operators.
\item Using a similar decoupling limit analysis we identify ghostly operators for non-Einstein-Hilbert kinetic interactions in the vielbein language in second-order form. In particular we consider the promising class of kinetic interactions proposed in \cite{Noller:2014ioa} as well as the Jordan frame of the linear effective vielbein matter coupling. As we will discuss, both were shown to be ghostly in the decoupling limit analysis of \cite{deRham:2013tfa}.
\end{itemize}
{\it Outline.} In Section \ref{sec:review} we will review the proposals that have been made for non-standard kinetic interactions and matter couplings. In Section \ref{sec:ADM}, we perform an ADM analysis to show that the most promising candidate modifications to the kinetic structure and matter couplings contain a BD ghost. In Section \ref{sec:matter-dl} we consider a general non-minimal matter coupling and will show using a standard decoupling limit analysis that the linear effective vielbein matter coupling is the unique choice that is ghost free at $\Lambda_3$. In Section \ref{sec:effective-vielbein-dl}, we focus on the linear effective vielbein coupling, and use the new decoupling limit to place an upper bound on the cutoff of this theory. In Section \ref{sec:kinetic-dl}, we consider kinetic interactions in second-order form in the vielbein language, and show that these suffer from ghosts as well.
\\

\noindent
{\it Conventions.} We will work with a mostly plus metric signature. We symmetrize and anti-symmetrize tensors with unit weight, so for example $T_{(\mu\nu)} \equiv \frac{1}{2} (T_{\mu\nu} + T_{\nu\mu})$. We use Greek letters $\mu,\nu,\cdots$ to represent tangent space indices and $a,b,\cdots$ to represent Local Lorentz indices. We denote the inverse of a vielbein $e^a_\mu$ by $e^\mu_a$. Whenever we work in the massive gravity limit, to emphasize that $f^a_\mu$ is a fixed reference vielbein, we will write $(f_{\rm ref})^a_\mu$.

\section{Review of extensions to ghost-free bi-gravity and massive gravity}
\label{sec:review}
\subsection{Ghost-free bi-gravity}

In this work we will focus on bi-gravity, by which we mean a theory of two interacting spin-2 fields, with no other gravitational degrees of freedom. We will consider these as effective field theories with some cutoff scale $\Lambda_{\rm c.o.}$. We will discuss the cutoff in more detail below, but suffice to say that the cutoff is no larger than the Planck scale $\Lambda_{\rm c.o.} \lesssim \mpl$. As a starting point, we will assume that an effective field theory for bi-gravity has the following properties\footnote{It may possible to weaken some of these assumptons, so this may be viewed as a starting point to fix ideas.}:
\begin{enumerate}
\item The action is Lorentz invariant.
\item There is a maximally symmetric vacuum, which we will take to Minkowski for this work. That is, $g_{\mu\nu}=a f_{\mu\nu}=b \eta_{\mu\nu}$ is a vacuum solution for constants $a$ and $b$.
\item Perturbatively around the Minkowski vacuum state, the theory describes a massless spin-2 field (with 2 degrees of freedom) and a massive spin-2 field (with 5 degrees of freedom). 
\item  We are able to trust the Minkowski vacuum solution within the regime of validity of the effective field theory. 
\end{enumerate}
In order for the Minkowski vacuum to be within the regime of validity of the effective field theory, all degrees of freedom should have non-vanishing kinetic terms about Minkowski space. Otherwise those degrees of freedom are infinitely strongly coupled around Minkowski, and we cannot trust the Minkowski solution within the regime of validity of the effective theory. Requirement 4 is automatically satisfied if we require that non-perturbatively, using an ADM analysis, bi-gravity contains seven degrees of freedom. Then assuming we have satisfied Requirement 3, all the degrees of freedom have non-vanishing kinetic terms around Minkowski. 
\\

Ghost-free bi-gravity satisfies all the requirements above. In the vielbein formalism, the action is given by \cite{deRham:2010kj,Hinterbichler:2012cn,Hassan:2011zd}
\be
\label{eq:bi-gravity}
S = M_e^2 S_{EH}[e] + M_f^2 S_{EH}[e] +m^2 M_{\rm pot}^2 S_{\rm pot}[e,f] +S_{\rm matt}[e,\psi_{\rm m}],
\ee
where $M_{\rm pot}^2 \equiv (M_e^{-2} + M_{f}^{-2})^{-1}$, and where $\psi_{\rm m}$ denotes a generic set of matter fields coupled covariantly to the vielbein $e$. Here we have defined the Einstein-Hilbert action for a vielbein $e^a_\mu$
\be
S_{EH}[e] \equiv \frac{1}{2} \int \d^4 x \ |e| R[e],
\ee
where $R[e]$ is the Ricci curvature scalar built out of $e$. As written, this action is in second-order form, in the sense that the connection is considered a function of the vielbein, rather than being an independent field. Explicitly, the connection is given by
\be
\om^{ab}_\mu =  e^c_\mu \left(O^{ab}_{\ \ \ c} - O_c^{\ \ ab} - O^{b\ \ a}_{\ c}\right),
\ee
where $O^{abc} \equiv e^{a \alpha} e^{b \beta} \partial_{[\alpha} e^c_{\beta]}$.
We also define the ghost-free potential interactions
\be
S_{{\rm pot}}[e,f] = -\frac{1}{8} \int \sum_{n=0}^{4} c_n\ \ep_{a_1 a_2 a_3 a_4} \prod_{i=1}^{n} e^{a_i} \prod_{j=n+1}^4 f^{a_j},
\ee
where wedge products between forms are implied, $e^a f^b \equiv e^a \wed f^b$. We choose the $c_n$ to ensure $e^a_\mu=\delta^a_\mu$ and $f^a_\mu=\delta^a_\mu$ is a vacuum solution, or equivalently that there are no tadpoles (which in these conventions amounts to $\sum_n c_n = \sum_n n c_n =0$). We will also demand $\sum_n n^2 c_n = -4$ so that around flat space the massive mode has a positive mass given by $m$. Otherwise we will leave the parameters arbitrary. 

Finally, we can add a matter sector which we denote as $S_{\rm matt}[e,\psi_{\rm m}]$ to denote a generic matter sector where the fields $\psi_{\rm m}$ are coupled to the vielbein $e$. We can also add a separate matter sector $S_{\rm matt}[f,\chi_j]$ where the fields $\chi_j$ were minimally coupled only to $f$, but there are no direct interactions between $\psi_{\rm m}$ and $\chi_j$.

In the rest of this section, we will work in the massive gravity limit, for which
\be
\label{eq:mg-limit}
M_f \rightarrow \infty, \ \ M_e \equiv \mpl \ \ {\rm (massive\ gravity\ limit).}
\ee
This decouples the canonically normalized fluctuations of $f$, so $f^a_\mu$ is fixed to its background value, $f^a_\mu \rightarrow (f_{\rm ref})^a_\mu$. If there is a matter sector coupled to $f$, this matter sector should be scaled as well. In this section we typically take the reference vielbein to be flat $(f_{\rm ref})^a_\mu = \delta^a_\mu$.

\subsection{Non-minimal matter couplings}
\label{sec:matter-coupling-review}
Since bi-gravity and massive gravity are theories with two metrics, it logically possible for a single matter sector to be simultaneously coupled to both vielbeins or metrics. The possibility was recognized by \cite{Hassan:2011zd}. There it was pointed out that a generic coupling to two metrics would likely ruin the constraint structure preventing the existence of the BD ghost, and would also likely be ruled out by tests of the equivalence principle. Nevertheless, there could still be special choices for the coupling to matter that could potentially preserve the constraint structure and lead to interesting phenomenology. 
\\

In \cite{Akrami:2013ffa}, the duality of the bi-gravity action \eqref{eq:bi-gravity} under the interchange of the two vielbeins $e\leftrightarrow f$ was used to motivate the study of matter couplings that would also preserve this duality. In \cite{Akrami:2014lja} it was argued that the couplings could be given a geometric interpretation in terms of a Finsler metric. Matter couplings to two vielbeins can also be motivated by dimensional deconstruction \cite{deRham:2013awa}. However, in \cite{Yamashita:2014fga}, an ADM analysis was performed which showed that generically when a matter sector is coupled to two sectors, a BD ghost is present. This result was also obtained using different methods in \cite{deRham:2014naa,Noller:2014sta}.
\\

There is a simple argument showing that we expect a general matter coupling obeying a weak equivalence principle to contain a ghost (see also \cite{Schmidt-May:2014xla} for a version of this argument).
\footnote{As we will dicuss in Section \ref{sec:matter-dl}, giving up the weak equivalence principle does not help, because we lose conservation of the stress energy tensor.} The most general matter coupling obeying the weak equivalence principle is given by
\be
S = M_e^2 S_{EH}[e] + M_f^2 S_{EH}[f] + m^2 M_{\rm pot}^2 S_{\rm pot}[e,f] + S_{\rm matt}[V(e,f),\psi_{\rm m}]
\ee
where $V(e,f)$ is an arbitrary, non-linear combination of the dynamical vielbeins $e$ and $f$. Inverting the relationship between $e$ and $V$ and sending $e\rightarrow e(V,f)$, we can write this as
\be
S = M_e^2 S_{EH}[e(V,f)] + M_f^2 S_{EH}[f] + m^2 M_{\rm pot}^2 S_{\rm pot}[e(V,f),f] + S_{\rm matt}[V,\psi_{\rm m}]
\ee
In this form, we see that we have modified the form of the potential interactions. As a result we expect the BD ghost to re-appear.
\\

In \cite{deRham:2014naa,Noller:2014sta} an interesting coupling was found that avoids this argument. This coupling does introduce new degrees of freedom  \cite{deRham:2014naa,Soloviev:2014eea,deRham:2014fha,deRham:2015cha}. However it was also shown in \cite{deRham:2014naa} that there is a decoupling limit in which no ghosts appear, and the ghost does not appear to quadratic order in perturbations around the mini-superspace. Thus, it was argued that this coupling could potentially be of interest phenomenologically.

This coupling, which we refer to as the linear effective vielbein coupling, is given by minimally coupling matter to the effective vielbein\footnote{We use a lower case $v$ to distinguish a linear combination from an arbitrary nonlinear combination, given by $V$}
\be
V^a_\mu = v^a_\mu \equiv \alpha e^a_\mu + \beta f^a_\mu.
\ee
This coupling evades the above argument because the ghost-free potential terms $S_{\rm pot}[e,f]$ are polynomials of the vielbeins, and so under this field redefinition a ghost-free potential is mapped to a ghost-free potential. If we impose the symmetric vielbein condition $\eta_{ab} e^a_\mu f^b_\nu = \eta_{ab} e^a_\nu f^b_\mu$, then this coupling can be described in terms of an effective metric
\be
g^{\rm eff}_{\mu\nu} = \alpha^2 g_{\mu\nu} + 2 \alpha \beta g_{\mu\alpha} \sqrt{g^{-1} f}^\alpha_{\ \ \nu} + \beta^2 f_{\mu\nu}.
\ee
However, the symmetric vielbein condition can be modified by the presence of this coupling, as discussed in \cite{deRham:2015cha,Hinterbichler:2015yaa}, so the metric and vielbein formulations are not equivalent unless the symmetric vielbein condition is added as a constraint.
\\

The linear effective vielbein coupling can also be derived by demanding that, at one loop, matter loops generate a ghost-free potential interaction. As a connection between modifications of the matter coupling and kinetic terms, we note that, in fact, at one loop the linear effective vielbein matter coupling will also generate a new kinetic interaction
\be
\label{eq:one-loop-kt}
\Gamma_{\rm 1-loop} \supset c_1 L \mpl^4 \int \d^4 x \sqrt{-g^{\rm eff}} +c_2  L \mpl^2 \int \d^4 x \sqrt{-g^{\rm eff}}R^{\rm eff} + O((R^{\rm eff})^2),
\ee
where $c_1$ and $c_2$ are dimensionless parameters and $L \equiv \log(\mpl/ \mu)$, where $\mu$ is a renormalization scale. The second term, $\sqrt{-g^{\rm eff}} R^{\rm eff}$, is a new kinetic interaction that was not present at tree level, but that appears at the same order in the curvature expansion as the normal Einstein-Hilbert term. We also note, as discussed in \cite{deRham:2014naa}, that the linear effective vielbein matter coupling spoils the possibility of the potential term being technically natural, as can be seen from the fact that the potential term is renormalized $\sim \mpl^4$.
\\

The requirement that, one loop, the matter interaction generates a ghost-free potential interaction was used by \cite{Heisenberg:2014rka} to propose a generalized version of the linear efective vielbein coupling
\be
\label{eq:one-loop-inspired-effective-vielbein}
V(e,f)^a_\mu \equiv \left(\frac{|e| + |f|}{|U(e,f)|}\right)^{1/4}U(e,f)^{a}_\mu,
\ee
where $U^a_\mu$ is a general (non-linear) combination of vielbeins. Based on the argument earlier in this section, we would expect this coupling to have a ghost because after a field redefinition we essentially have changed the form of the ghost-free potential. We will verify this below.
\\

As this work was being completed, in \cite{Huang:2015yga,Heisenberg:2015iqa} it was shown that the only matter coupling that does not give rise to ghosts in the decoupling limit is given by the effective vielbein coupling. The results of this work are in complete agreement with what is found there, but is complementary in the sense that we make different assumptions about the matter couplings we consider (we do not assume a weak equivalence principle). Furthermore, we will extend that work by considering a more different decoupling limit in which the interactions of the linear effective vielbein coupling do not vanish, and we put an upper bound on the cutoff scale for this coupling.

\subsection{New kinetic interactions}
In a related but separate development, the possibility of a ghost-free kinetic interaction for a massive graviton beside Einstein-Hilbert has been considered. We will consider a kinetic interaction to be a generalization of the Einstein-Hilbert term in that it contains two derivatives in unitary gauge, though it can be non-linear in $e$ and $f$. Perturbatively around Minkowski space, the kinetc term for a massless spin-2 field $H_{\mu\nu}$ is fixed (up to field redefinitions) to given by the Fierz-Pauli kinetic term \cite{Fierz:1939ix}
\be
\mathcal{L}_{\rm FP} = \ep^{\mu\nu\rho\sigma} \ep^{\mu'\nu'\rho'\sigma'} \partial_\mu H_{\nu\nu'} \partial_{\mu'} H_{\rho\rho'} \eta_{\sigma \sigma'}.
\ee
In \cite{Hinterbichler:2013eza} (see also \cite{Folkerts:2011ev}), it was shown that, for a massive spin-2 field, in addition to the Fierz-Pauli kinetic term there are also ghost-free kinetic interactions $\sim \partial^2 H^n$ for $n>2$. In four dimensions there is one such interaction, of the form
\be
\label{eq:new-kinetic-cubic}
\mathcal{L}_{\rm der} = \ep^{\mu \nu \rho \sigma} \ep^{\mu' \nu' \rho' \sigma'} \partial_{\mu} H_{\nu\nu'} \partial_{\mu'} H_{\rho \rho'} H_{\sigma \sigma'}.
\ee
The ghost-freedom can be verified, for example, in the ADM formalism. Once we couple $H_{\mu\nu}$ to matter, by standard arguments \cite{Feynman:1996kb} we will inevitably be forced to add futher non-linear interactions for $H$. Thus it is important to ask if there is a non-linear version of Equation \eqref{eq:new-kinetic-cubic} that propagates the same number of degrees of freedom non-linearly.
\\

The first issue we face when modifying the kinetic term beyond the linear level is that we have to choose how to describe the spin-2 field non-linearly. We can identify several different possible approaches that can be taken in constructing an extension to the kinetic term:
\begin{itemize}
\item {\it Metric vs. Vielbein Language:} Do we choose to write the theory in metric or vielbein variables?
\item {\it First-order form vs. Second-order form:} Do we treat the connection associated with the metric or vielbein as an independent field? 
\end{itemize}
For ghost-free massive gravity with an Einstein-Hilbert term and minimal matter couplings, the four different formulations are all equivalent.\footnote{Strictly speaking in the vielbein there are additional branches of the theory that are not equivalent to the metric language \cite{Banados:2013fda}, however in this work we will restrict ourselves to the branch of the theory where the BD ghost is not present.} For modified kinetic interactions, however, there are (at least) four different approaches to constructing a non-linear action, which may lead to different theories.
\\

In \cite{deRham:2013tfa}, a no-go theorem was given for massive gravity in the metric language in four spacetime dimensions in second-order form which ruled out the possiblity of ghost-free non-linear kinetic interactions besides the Einstein-Hilbert term. This argument is reviewed in Appendix \ref{appendix}. While the argument was given for massive gravity, the argument also applies to bi-gravity,  as massive gravity is a limit $M_f \rightarrow \infty$ of bi-gravity. Thus this result shows that for bi-gravity theories, there are no kinetic interactions, at least in the metric language.
\\

Modifications of the kinetic term in first-order form in the vielbein language, partly motivated by dimensional deconstruction of the Gauss-Bonnet term \cite{Lovelock:1971yv} in five dimensions, were considered in \cite{deRham:2013awa}. There it was shown that new kinetic interactions give rise to new degrees of freedom in the decoupling limit. The Lorentz \stu fields, which are non-dynamical in massive gravity and bi-gravity \cite{Ondo:2013wka}, can become dynamical once the Lorentz invariance of the Einstein-Hilbert structure is broken, leading to new degrees of freedom.
\\

In \cite{Noller:2014ioa}, it was argued that there could be kinetic interactions that are ghost-free in the decoupling limit, even if there is a ghost non-linearly. As an example, the linear effective vielbein matter coupling can also be described in Jordan frame, where it appears as a new kinetic interaction. Starting from
\be
S = M_e^2 S_{EH}[e]+ M_f^2 S_{EH}[f] + m^2 M_{\rm pot}^2 S_{\rm pot}[e,f] + S_{\rm matt}[v(e,f),\psi_{\rm m}],
\ee
and inverting the relationship between $e$ and $v$ by sending $e \rightarrow e(v,f)$ we arrive at the action
\be
S= M_e^2 S_{EH}[\alpha^{-1}(v - \beta f)] + M_f^2 S_{EH}[f] +m^2 M_{\rm pot}^2 \hat{S}_{\rm pot}[v,f] + S_{\rm matt}[v,\psi_{\rm m}],
\ee
where now the fundamental fields that we vary in the action are $f$ and $v$ rather than $f$ and $e$. The notation $\hat{S}_{\rm pot}[v,f]$ indicates that the parameters of the potential term change under this field redefinition. It was argued that since the linear vielbein coupling is ghost free in the decoupling limit, this kinetic interaction should be as well. We will revisit this point in Section \ref{sec:kinetic-dl}, given the new decoupling limit we use for the matter coupling.

Nevertheless, based on this argument, a generalized set of kinetic interactions was proposed in the vielbein formalism in second-order form that could also be ghost-free in the decoupling limit
\be
\label{eq:2of-nkt}
S = M_e^2 S_{EH}[e] + M_v^2 S_{EH}[\alpha e+\beta f] + M_f^2 S_{EH}[f] + m^2 M_{\rm pot}^2 S_{\rm pot}[e,f] + S_{\rm matt}[e,\psi_{\rm m}].
\ee
We note that this interaction can be interpreted as the one loop kinetic term generated by the linear effective vielbein coupling given in Equation \eqref{eq:one-loop-kt}. This interaction was considered for Dark Matter in \cite{Blanchet:2015sra}.
\\

In this work we will close the gap and show that in the vielbein language in second-order form, there is still no loophole to the metric argument. We will consider the kinetic term, and show that there is a decoupling limit where the ghost appears at $\Lambda_3$. In fact, as we will discuss in Section \ref{sec:kinetic-dl}, this kinetic term was already shown to contain a ghost in this decoupling limit by the no-go analysis of \cite{deRham:2013tfa}.

\subsection{The possibility of an effective field theory description}
\label{sec:possibility-of-eft-description}
As discussed in the beginning of this section, we take the point of view that bi-gravity is an effective field theory that breaks down at some cutoff scale $\Lambda_{\rm c.o.}$. Of course in gravity we always imagine that the cutoff is no higher than $\mpl$, but in principle this scale could also be lower. For example, $\Lambda_{\rm c.o.}$ could be the mass of new degrees of freedom that are necessary to make the theory consistent. From this perspective, it could be that asking that theory has five degrees of freedom at all energy scales or that the theory has five degrees of freedom doing a full ADM analysis gives too strong a requirement.
\\

In order for the theory to describe a Vainshtein mechanism, the cutoff should be larger than $\Lambda_3$. This is because, as discussed in the introduction, the Vainshtein mechanism relies on operators at the scale $\Lambda_3$ becoming large. So if we want our effective theory to have a Vainshtein mechanism within the regime of validity of the theory, we must have $\Lambda_{\rm c.o.} > \Lambda_3$. In the language of \cite{deRham:2014wfa}, $\Lambda_3$ is a strong coupling scale, but is not the cutoff of the effective field theory.

In ghost-free massive gravity and bi-gravity, ghost-freedom in the ADM sense means that it may be consistent to consider that the cutoff larger is than $\Lambda_3$, since it is possible that full unitarity is maintained above this scale. 
\\

However, for the new kinetic interactions and matter couplings, there typically are ghostly operators that arise at some scale $\Lambda_{\rm ghost}$\footnote{More accurately they arrange at a range of scales $\Lambda_{{\rm ghost},n}$, we will focus here on the lowest scale.}. Since these operators break full (not just perturbative) unitarity, we can safely infer that there must be a cutoff for these theories $\Lambda_{\rm c.o.} < \Lambda_{\rm ghost}$, at which the effective field theory description must break down. It is also possible for unitarity to be broken by some other mechanism other than the appearance of a ghost, however in this work will focus on ghosts.
\\

Therefore it is interesting to ask if it is possible for the new kinetic interactions or matter couplings to have a strong coupling regime, in which we could trust the strongly coupled interactions without being sensitive to an unknown UV completion. In the language of \cite{deRham:2014wfa}, we can ask if there are irrelevant but important interactions coming from the kinetic interaction or matter coupling. As a minimum requirement to be able to trust a given operator in a strong coupling regime, we should ask for that operator to not introduce a ghost. As we will see, even this minimal requirement is very restrictive. A generic modification of the matter coupling or kinetic structure breaks unitarity at the lowest possible scale because the pure scalar or scalar-matter interactions are ghostly.

In \cite{deRham:2014naa}, it was shown that the linear effective vielbein coupling is special, in that the pure scalar-matter interactions arising from the linear effective vielbein matter coupling gave rise to second order equations of motion for the helicity-0 mode of the graviton. As a result, it was argued that the scalar-matter interactions could be treated as irrelevant but important interactions coming from the linear effective vielbein matter coupling. 
\\

Thus, it was argued in \cite{deRham:2014naa,Huang:2015yga,Heisenberg:2015iqa} that we can potentially trust the predictions, including a strongly coupled regime, 
of an extended matter coupling (or, as argued in \cite{Noller:2014ioa}, of a new kinetic term) if we can work in a regime where $\Lambda_3 \ll \Lambda_{\rm c.o.} < \Lambda_{\rm ghost}$. In this picture $\Lambda_{\rm c.o.}$ is an unknown scale that is not predicted within the effective field theory, other than being smaller than $\Lambda_{\rm ghost}$\footnote{This is similar to how the Higgs mass is not predicted with an effective field theory containing massive $W$ and $Z$ bosons, even though there is a calculable scale $\sim \rm{TeV}$ at which (perturbative) unitarity is broken in that effective theory.}. From this perspective, it becomes important not just to determine whether or not one or more ghosts is present, but also to establish the lowest scale at which ghostly operators arise.

\section{ADM analysis in the constrained tri-gravity picture}
\label{sec:ADM}
In this section, we consider the most promising proposals for extensions of the matter coupling and kinetic interactions in the vielbein language in second-order form. We show that these extensions introduce new {\it d.o.f.s} by performing an ADM analysis.

While at first glance the new kinetic term in Equation \eqref{eq:2of-nkt} may look related to a bi-gravity theory, we will find it more useful to view it as a constrained tri-gravity theory. We can write the new kinetic interactions in four dimensions as
\ba
\label{eq:constrained-trigravity-nkt}
S &=& M_e^2 S_{EH}[e] + M_f^2 S_{EH}[f] +  M_{v}^2 S_{EH}[v]  +  m^2 M_{\rm pot}^2  S_{\rm pot}[e,f]  \nn \\
&&+ \int \d^4 x \ \lambda^\mu_a \left(v^a_\mu - \alpha e^a_\mu - \beta f^a_\mu \right),
\ea
where $\lambda^\mu_a$ are 16 Lagrange multiplier fields. Without the constraint, this theory would simply be bi-gravity plus a decoupled copy of GR. The presence of the constraint completely modifies the Hamiltonian structure, and reintroduces a ghost.

Therefore, without the constraint, we have three copies of (diagonal) diffeomorphisms (diffs) and Local Lorentz transformations (LLTs). The constraint breaks this down to a single copy of diffs and LLTs (if we allow $\lambda$ to transform appropriately). In the massive gravity limit $M_f \rightarrow \infty$ and $f$ is fixed to its background value, then these diagonal copies are broken and there are no gauge symmetries. 

Taking the massive gravity limit (given in Equation \eqref{eq:mg-limit}) of the action \eqref{eq:constrained-trigravity-nkt}, we find a modified massive gravity action
\be
\label{eq:mg-nkt}
S = M_e^2 S_{EH}[e] + M_v^2 S_{EH}[v] + m^2 M_{\rm pot}^2 S_{\rm pot}[e,f_{\rm ref}] + \int \d^4 x \lambda^\mu_a \left(v^a_\mu - \alpha e^a_\mu - \beta (f_{\rm ref})^a_\mu \right).
\ee
We will directly perform an ADM analysis on the massive gravity limit and find that there is a ghost. This might be surprising because at first sight it might have appeared that the action in Equation \eqref{eq:mg-nkt} could arise as a limit of bi-gravity. However this is not possible. Instead we see that it is best viewed as a limit of a constrained tri-gravity theory.

We can also view the matter coupling from this perspective
\ba
S &=& M_e^2 S_{EH}[e] + M_f^2 S_{EH}[f] + m^2 M_{\rm pot}^2 S_{\rm pot}[e,f_{\rm ref}] + S_{\rm matt}[v,\psi_{\rm m}] \nn \\
&& \int \d^4 x \lambda^\mu_a \left(v^a_\mu - \alpha e^a_\mu - \beta (f_{\rm ref})^a_\mu \right).
\ea
We will outline the ADM analysis of the matter coupling in Section \ref{sec:ADM-matter}. The matter couplings were shown to have a ghost in the vielbein formulation in \cite{deRham:2015cha}. Our formalism gives a different perspective on how the constraint structure is modified, but the conclusion is the same. 

\subsection{Hamiltonian analysis of new kinetic term}
In this section we focus on the massive gravity limit defined by Equation \eqref{eq:mg-limit} where $f$ is fixed. It is easy to extend the analysis below to allow $f$ to be dynamical. 

A very useful consequence of writing things in the constrained form of Equation \eqref{eq:constrained-trigravity-nkt} is that the distinction between first and second-order form vanishes, provided that we do not subsitute the solution to the $\lambda$ constraints back into the action. To see this, note that the equations of motion for the connections related to $e$ and $v$ are just the normal torsion-free conditions. Working in first-order form simplifies the analysis in $D=3$. 

\subsubsection{$D=3$ case}

Since the potential term is not crucial to the analysis, we will not include it (it does not change any of the counting arguments below). We will also set $M_e = M_v = 1$ in this section since it is not necessary to track these scales for the ADM argument. The action is 
\be
S = \int \ep_{abc} \left( R[\om]^{ab}  e^c + R[\mu]^{ab}  v^c \right)  + \int \d^3 x \lambda^\mu_a \left(v^a_\mu - \alpha e^a_\mu - \beta (f_{\rm ref})^a_\mu \right),
\ee
where $f^a_\mu$ is some fixed tensor that we do not vary, and the Riemann two form is defined by
\be
R[\om]^{ab} \equiv \d \om^{ab} + \om^{ac} \om_c^{\ b}.
\ee
Passing to the Hamiltonian we find
\ba
\mathcal{H} &=& e^a_0 P_a[\om] + \om^{ab}_0 M_{ab}[e,\om]  + v^a_0 P^a[\mu] + \mu^{ab}_0 M^{ab}[v,\mu] + \lambda^\mu_a C^a_\mu[v,e,f_{\rm ref}],
\ea
where
\ba
P_a[\om] &\equiv& \ep_{abc}\ep^{ij} R_{ij}^{bc}[\om] = \ep_{abc} \ep^{ij}\left( \partial_i \om_j^{bc} + \om_i^{bd} \om_j^{dc} \right), \nn \\
M_{ab}[e,\om] &\equiv & \ep^{ab}_{\ \ \ c}\ep^{ij} T_{ij}^c[e,\om] = \ep^{ab}_{\ \ \ c} \ep^{ij} \left( \partial_i e_j^c + \om^{cd}_i e^d_j \right), \nn \\
C^a_\mu &\equiv& v^a_\mu - \alpha e^a_\mu - \beta (f_{\rm ref})^a_\mu.
\ea
The momentum conjugate to $e^a_i$ is
\be
\dot{e}^a_i \pi^i_a \implies \pi^i_a = -\ep_{abc}\ep^{ij} \om^{ab}_j.
\ee
Using the canonical Poisson bracket between $e$ and the conjugate momentum $\pi$ we can derive the relationship
\be
\{e^a_i, \om^{bc}_j\} = -\frac{1}{2} \ep_{ij} \ep^{abc}.
\ee
Thus we identify several non-dynamical equations\footnote{We distinguish between `constraints,' which are relationships between dynamical variables, and `non-dynamical equations,' which allow for the possibility that the equation can be solved for an auxiliary field like the lapse $e^0_0$ instead of for a dynamical variable.} (that is, equations that come from varying with a field that does not have a kinetic term)
\ba
P_a[\om] - \alpha \lambda^0_a &\approx& 0, \nn \\
P_a[\mu] + \lambda^0_a  &\approx& 0, \nn \\
M_{ab}[e,\om] &\approx& 0, \nn \\
M_{ab}[v,\mu] &\approx& 0, \nn \\
v^a_\mu - \alpha e^a_\mu - \beta (f_{\rm ref})^a_\mu  &\approx& 0.
\ea
Taking the Poisson bracket of the above conditions with the Hamiltonian we obtain further non-dynamical equations
\ba
2 \om^{a}_{0,b} P^b[\om] - \alpha \ep^{ij} D_i[\om] \lambda^a_j  &\approx& 0, \nn \\
2 \mu^{a}_{0,b} P^b[\mu] + \ep^{ij} D_i[\mu]\lambda^a_j  &\approx& 0, \nn \\
-2 e^{[a}_0 P^{b]}[\om] - \alpha \lambda^{[a}_i e^{b]}_i  &\approx& 0, \nn \\
-2 v^{[a}_0 P^{b]}[\mu] + \lambda^{[a}_i v^{b]}_i  &\approx& 0, \nn \\
D_k[\mu] v^a_0 - \alpha D_k[\om] e^a_0 - \mu^{a}_{0,b} v^b_k + \alpha \om^{a}_{0,b} e^b_k  &\approx& 0,
\ea
where $D_k[\om]$ is the Lorentz covariant derivative using the spin connection $\om$.

Not all of the $C^a_\mu$ have generated further non-dynamical equations. The reason is that there is no momentum conjugate to $e^a_0$, so $C^a_0$ trivially commutes with the Hamiltonian.
\\

It is now useful to perform a counting argument. In three dimensions, the phase space contains 45 fields\footnote{Technically there are 54 fields, if we include the momenta conjugate to the $\lambda^a_\mu$. This makes sense because the phase space should be even dimensional. However, these momenta will be immediately removed with 9 secondary constraints, so we ignore these fields.}
\be
9\ \lambda^a_\mu+ 6 \ e^a_i+ 6\ \om^{ab}_i+  6\ v^a_i+  6\ \mu^{ab}_i+ 3\ e^a_0+ 3\ \om^{ab}_0+ 3\ v^a_0+ 3\ \mu^{ab}_0 = 45 \ {\rm fields}.
\ee
Meanwhile, there are 39 non-dynamical equations. 

Then, assuming we can solve all of the non-dynamical equations for all of the auxiliary fields (we can do this perturbatively), we are left with a phase space
\be
\rm{dimension\ of\ phase\ space} = 45 - 39 = 6.
\ee
A healthy massive graviton in $D=3$ has 4 phase space degrees of freedom, so this counting is consistent with the existence of a BD ghost.

In ghost-free bi-gravity, the equation for the lapse $e^0_0$ is a constraint after integrating out the shift, and the auxiliary components of the vielbein \cite{Hinterbichler:2012cn}. This structure is broken by the $\lambda$ constraint, however. After integrating out all of the auxiliary fields, the equation for the lapse $e^0_0$ can be solved for the lapse, rather than being a constraint. This occurs because the equation that determines $\mu$ involves the lapse and the shift in a highly non-linear way. We will demonstrate that this occurs in the mini-superspace below.

If all of these equations can be solved for the auxiliary variables, then there are no constraints. Commuting these conditions with the Hamiltonian then generate equations for the time derivatives of the auxiliary variables, there are no constraints. In other words, if we can solve for the lapse
\be
N = N(e^a_i, \om^{ab}_i).
\ee
Then the Poisson bracket of $N$ with the Hamiltonian will simply give us an equation for $\dot{N}$, it will not give us an independent constraint on the dynamical variables $e^a_i$ and $\om^{ab}_i$. In other words, the Dirac procedure terminates with no constraint to remove the BD ghost.

\subsubsection{Analysis for $D$ dimensions}

\emph{Review of Massive Gravity.}

First it is useful to review the counting for massive gravity in $D$ dimensions in first-order form. We will consider the generic case where the BD ghost is present.

There are
\be
D^2\ e^a_\mu + \ \frac{D^2(D-1)}{2}\ \om^{ab}_\mu = \frac{D^2(D+1)}{2} \ {\rm fields}.
\ee
There are no first class constraints because diffs and LLTs are explicitly broken by the reference vielbein.

There are $D$ equations of motion coming from $e^a_0$. When a BD ghost is present, these are solved for the lapse and shift and there is no constraint among these equations. 

Then there are $\frac{D(D-1)}{2}$ equations coming from varying $\om^{ab}_0$. These equations are constraints. One interpretation is that these constraints guarantee that the theory is equivalent to a metric formulation (which ultimately requires that $\frac{D(D-1)}{2}$ components of the vielbein remain unphysical). If these equations fail to be constraints, there are additional {\it d.o.f.s} in the phase space that we expect to be ghostly (even if they are not ghostly, they correspond to {\it d.o.f.s} that are infinitely strongly coupled around the Minkowski vacuum). We further expect these primary constraints to generate $\frac{D(D-1)}{2}$ secondary constraints. The time derivatives of the secondary constraints will result in $\frac{D(D-1)}{2}$ equations that can be solved for the $\om^{ab}_0$ themselves.

Lastly, there are second class constraints that remove some redundant components of the spin connection. In $D$ dimensions, there are $D(D-1)$ spatial components of the vielbein $e^a_i$, all of which have kinetic terms. However there are $\frac{D(D-1)^2}{2}$ spatial components of the spin connection $\om^{ab}_i$, which means that some of the spin connection components are redundant and are not momenta conjugate to the vielbeins. There are $\frac{1}{2}D(D-1)(D-3)$ non-dynamical equations that can be solved for these redundant vielbein components.

To summarize, for massive gravity, there are:
\be
D+\frac{3}{2}D(D-1) + \frac{D(D-1)(D-3)}{2}= \frac{1}{2}D(D^2-D+2) \ \rm {non-dynamical\ equations}.
\ee
Thus the total phase space is
\ba
{\rm phase\ space\ dimension}&=&\frac{D^2(D+1)}{2}\ {\rm fields} - \frac{1}{2}D(D^2-D+2)\ {\rm n.d.\ equations} \nn \\
&=& 2\left[\frac{D^2 - D - 2}{2}  + 1\right] \ {\rm {\it d.o.f.s}}.
\ea
This is the right amount for a massive graviton in $D$ dimensions + 1 BD ghost. Ghost-free massive gravity contains two additional secondary constraints that remove the BD ghost.
\\ 
\\
\emph{New Kinetic Interaction}.

The new kinetic interaction counting is now relatively easy to perform. Because of the broken diffs, we expect the counting for both $e$ and $v$ separately to follow the generic massive gravity counting above before imposing the constraints associated with $\lambda$. Namely we expect
\be
{\rm fields\ in\ } e,v,\om,\mu - {\rm n.d.\ equations} = 2 D(D-1).
\ee
Next we impose the constraints from $\lambda^a_\mu$. There are $D^2$ primary constraints. However as we have seen above, only $D(D-1)$ of these lead to secondary constraints. Thus
\be
{\rm dimension\ of\ phase\ space} = 2D(D-1) + D^2 - D^2 - D(D-1) = 2\left[\frac{D^2-D-2}{2}+1\right].
\ee

In order to remove the BD ghost, we would need that, after integrating out all of the nondynamical fields (that is, $\lambda,\mu,v$ and the unphysical parts of $e,\om$), there was a constraint equation for $e$ and $\om$. However given that this does not occur for $D=3$ as we have seen above, it cannot occur for $D>3$. We will also show this explicitly in the mini-superspace.

\subsubsection{Mini-superspace}
To illustrate these points, we demonstrate how the above procedure works in the mini-superspace. This argument will hold in $D$ dimensions.

In the mini-superspace we write $\lambda^0_0 = L$ and $\lambda^i_j = \lambda \delta^i_j$. We also define lapses $e^0_0 = N$, $v^0_0 = K$ and scale factors $e^i_j = a \delta^i_j, v^i_j = b \delta^i_j$ and the spin connections $\om^{i0}_j = \om\delta^i_j, \ \mu^{i0}_j = \mu \delta^i_j$ (the other components vanish). We will also introduce a matter source given by a perfect fluid with vanishing pressure that is minimally coupled to $a$. Then the action becomes
\ba
S &=& \int \d^D x \  \Big[ a^{D-3} (D-2)\(\omega \dot{a} - \omega^2 N\) +b^{D-3} (D-2) \(\mu \dot{b} - \mu^2 K\) \nn \\
&& + L \left(K - \alpha N - \beta\right) + \lambda\left(b - \alpha e - \beta\right) + \mathcal{L}_{\rm matt}\Big].
\ea
The Hamiltonian is
\be
\mathcal{H} = P_a^2 N + P_b^2 K + L(K - \alpha N - \beta) + \lambda(b - \alpha a - \beta) - N a^{D-3} \rho(a).
\ee
where the momenta $P_a$ and $P_b$ are given by
\be
P_a \equiv  a^{D-2} (D-2) \om, \ \ P_b \equiv b^{D-2}(D-2) \mu.
\ee
There are four non-dynamical equations that follow immediately
\ba
C_1 &=& K - \alpha N - \beta \approx 0, \nn \\
C_2 &=& b - \alpha a - \beta \approx 0, \nn \\
C_3 &=& P_a^2 - \alpha L -  a^{D-3} \rho \approx 0, \nn \\
C_4 &=& P_b^2 + L \approx 0.
\ea
The time derivative of $C_2$ generates another non-dynamical equation\footnote{Commuting $C_3$ and $C_4$ with $H$ just generates the condition $\lambda=0$.}
\ba
C_5 &\equiv& \{C_2, H\} = K P_b - \alpha N P_a \approx 0.
\ea
Solving $C_5$ for $P_b$ and $C_1$ for $K$, we find that
\be
P_b = \frac{\alpha N P_a}{K} = \frac{\alpha N}{\alpha N + \beta} P_a.
\ee
The key point is that this is non-linear in the lapse. In GR (and in bi-gravity), the equation $C_3$ is a constraint on $P_a$. However, it now becomes an equation for the lapse
\be
C_3 = P_a^2 + \alpha P_b^2 = \left( 1 + \alpha^3 \left(\frac{N}{\alpha N + \beta} \right)^2\right) P_a^2 - a^{D-3} \rho \approx 0.
\ee
Unlike the ordinary Friedmann equation, this is an equation that determines $N$ in terms of $\rho$, rather than an equation for the scale factor. 

To summarize, we have demonstrated how the procedure outlined in the previous sections can be carried out in the mini-superspace, showing the existence of a ghost.

\subsection{Constrained vielbein picture for linear effective vielbein coupling}
\label{sec:ADM-matter}
The linear effective vielbein matter coupling was shown to lead to ghosts in the vielbein language in \cite{deRham:2015cha}. We outline an alternative argument demonstrating out current method to see that the matter coupling leads to new degrees of freedom in $D=3$ (and thus in higher dimensions as well. Our starting point is
\be
S = M_e^2 S_{EH}[e] + m^2 M_{\rm pot}^2 S_{\rm pot}[e,f_{\rm ref}] + S_{\rm matt}[v,\psi_{\rm m}] + \int \d^3 x\ \lambda^\mu_a (v^a_\mu - \alpha e^a_\mu - \beta (f_{\rm ref})^a_\mu).
\ee
As in the previous section, we will ignore the mass term (taking $m\rightarrow 0$) as it does not affect the conclusions that we will draw, and we will take $M_e = M_v = 1$. Then the action is
\be
S = S_{EH}[e] + S_{\rm matt}[v,\psi_{\rm m}] + \int \d^3 x\ \lambda^\mu_a (v^a_\mu - \alpha e^a_\mu - \beta (f_{\rm ref})^a_\mu).
\ee
We obtain non-dynamical equations
\ba
P_a - \alpha \lambda^0_a &\approx& 0, \nn \\
M^{ab} &\approx& 0, \nn \\
v^a_\mu - \alpha e^a_\mu - \beta (f_{\rm ref})^a_\mu &\approx& 0, \nn \\
T^{\mu\nu} v_\nu^a + \lambda^\mu_a &\approx& 0.
\ea
We can first solve the last equation for $\lambda^\mu_a$. Then the first non-dynamical equation $P^a \approx \alpha \lambda^a_0$ will involve the matter sector in a highly non-trival way, and this equation can be solved for the lapse, so it is not a genuine constraint.

The time derivatives of these generate further consistency conditions. Note that very conveniently $[P_{\rm matt},H]=0$ since $v$ has no conjugate momentum, so that the matter does not contribute to the consistency conditions in this language.
\ba
2 \om^{a}_{0,b} P^b - \alpha \ep^{ij} D_i [\om] \lambda^a_j &\approx& 0,\ \nn \\
-2 e^{[a}_0 P^{b]}[\om] - \alpha \lambda^{[a}_i e^{b]}_i &\approx& 0, \nn \\
\alpha\left( - D_k[\om] e^a_0 + \alpha \om^{a}_{0,b} e^b_k \right) &\approx& 0.
\ea

There are
\be
9\ \lambda^\mu_a + 9\ v^a_\mu + 3 e_0^a + 3\om_0^a = 24\ {\rm non-dynamical\ fields}.
\ee
We expect these to be determined by the non-dynamical equations. In addition there are
\be
6 e^a_i + 6 \om^{ab}_i = 12\ {\rm dynamical\ fields}.
\ee
This leads to a counting
\be
15 + 9\ {\rm primary\ conditions} + 12\ {\rm secondary\ conditions} = 27 + 9 = 36\ {\rm conditions}.
\ee
We expect that we can solve 24 of these for the 24 auxiliary variables. Then the remaining 3 are genuine constraints on the auxiliary variables. If we are optimistic, these 3 constraints will generate 3 secondary constraints, so there will be 6 constraints. That leads to 6 phase space {\it d.o.f.s}, corresponding to graviton plus a BD ghost in the best case scenario. 

\section{Cutoff for general non-minimal matter couplings}
\label{sec:matter-dl}
We have shown in the vielbein language how two of the most promising candidates for extensions of the kinetic term and matter coupling modify the constraint structure of ghost-free massive gravity so that the BD ghost returns. Based on the arguments from Section \ref{sec:possibility-of-eft-description} however, it may still be interesting to consider these extensions as effective field theories below some cutoff. The next three sections will address this question for a range of matter couplings and kinetic terms, including the linear effective vielbein coupling proposed by \cite{deRham:2014naa}.

In this section we show that a generic modified matter coupling will introduce new degrees of freedom at the scale $\Lambda_3$. We will consider a broad class of matter couplings, and we will find that the only coupling that is ghost-free in the decoupling limit for a generic matter sector is for the matter to be coupled to a single linear effective vielbein, as proposed by \cite{deRham:2014naa}. There are other possibilities if we allow for restrictions on the choice of matter sector and couple that sector in a specific way. For example we can have a scalar field whose kinetic term is coupled to a single linear effective vielbein and with an arbitrary number of potential terms coupled to different effective vielbeins (confirming the results of \cite{Yamashita:2014fga}). 

We consdier a generic matter coupling which is built out of an arbitrary number of ``effective vielbeins'' $V[e,f]^a_\mu$ (which we allow to be non-linear functions of $e$ and $f$, but not of their derivatives)
\be
\label{eq:matter-couplings}
S = M_e^2 S_{EH}[e] + M_f^2 S_{EH}[f] +  m^2 M_{\rm pot}^2 S_{\rm pot}[e,f] + \sum_V S_{\rm matt}^{(V)}[V,\psi_{\rm m}].
\ee
We do not assume that $S_{\rm matt}^{(V)}[V,\psi]$ represents a minimal coupling between $V$ and $\psi_{\rm m}$, and a given matter field may couple to multiple vielbeins $V$. In other words, we do not assume a weak equivalence principle. As an example of what we have in mind, consider a vector field $A_\mu$ coupled to two metrics
\be
S_{\rm matt} = \int \d^4 x |e| \left(-\frac{1}{4} g^{\mu\rho}_{(e)} g^{\nu\sigma}_{(e)}F_{\mu\nu} F_{\rho \sigma} \right) + \int \d^4 x |f| \left(- \frac{m_f^2}{2} g_{(f)}^{\mu\nu} A_\mu A_\nu  \right),
\ee
where $F_{\mu\nu} \equiv \partial_\mu A_\nu - \partial_\nu A_\mu$. We will consider this coupling explicitly below and show that in fact it contains a ghost in the decoupling limit. We note that there can be matter couplings not of form we are considering because within each term the matter may only couple to a single vielbein. For example our assumption does not include the quasidilaton \cite{D'Amico:2012zv} and extended quasidilaton extensions \cite{DeFelice:2013tsa}.

Our goal is not to perform an exhaustive search of all possible matter couplings. A search was performed in \cite{Huang:2015yga} assuming the weak equivalence principle, and ruled out any coupling that is ghost-free in the decoupling limit besides the linear effective vielbein coupling. Our goal is to identify the sources of ghosts in the decoupling limit. We will see how the known cases that are ghost-free evade these problems, and we will also see how several candidate extensions go wrong. However, we do not claim that we have found the most general matter coupling that is ghost free in the decoupling limit.\footnote{There are stronger requirements if one further demands that the graviton potential remains ghost-free after taking into account one loop corrections. For example, a scalar field with a kinetic term coupled to one metric and potential term coupled to another metric is ghost-free classically, but at one loop will detune the graviton potential.}
\\

We expect a generic matter sector to introduce new degrees of freedom even in the decoupling limit, based on the argument in Section \ref{sec:matter-coupling-review}. To reiterate, if the weak equivalence principle is satisfied we can always do a field redefinition to make the coupling to matter a standard one. For a general matter coupling, this field redefinition will dramatically alter the form of the potential term, and in that description it will be clear that a ghost is present in the decoupling limit. This problem only gets worse if the weak equivalence principle is violated. Nevertheless it is interesting to see exactly where the problem lies in the matter coupling. 
\\

As we will show below, the vielbein and metric formalisms are equivalent in the decoupling limit for matter couplings. In the metric language we can quickly see why we would expect a problem to arise. In the decoupling limit that we consider in this section, we will see that the metric and vielbein languages are equivalent. We can expand a generic effective metric as $g^{\rm eff}_{\mu\nu} = \eta_{\mu\nu} + c_1 \Pi_{\mu\nu} + c_2 \Pi^2_{\mu\nu} +\cdots$, where $c_{1,2}$ are constants and where $\Pi_{\mu\nu}\equiv \Lambda_3^{-3} \partial_\mu \partial_\nu \pi$. Then the matter Lagrangian will contain the term
\be
\label{eq:pi-matter-interaction}
\mathcal{L}_{\rm matt} \supset \frac{c_2}{\Lambda_3^6} \partial_\mu \partial_\alpha \pi \partial^\alpha \partial_\nu \pi T_{(\eta)}^{\mu\nu},
\ee
where $T^{\mu\nu}_{(\eta)}$ is the matter stress energy tensor on Minkowski space. There are also similar terms $\sim \partial^4 \pi^2 T$ that arise from evaluating $T^{\mu\nu}$ on the non-trivial background of $\pi$. Only special choices of the coupling could allow for second order equations of motion, given the inevitable appearance of higher derivatives on $\pi$ in the action. Furthermore, since in the decoupling limit only the helicity-0 mode $\pi$ is coupled to matter, not the tensors $h_{\mu\nu}$ or vectors $B_\mu$, there is no field redefinition that can remove these higher derivative terms.

\subsection{Equations of motion}
Our approach to taking is to first compute the equations of motion for the \stu fields exactly, then take the decoupling limit. We will use superscript $^{(V)}$ or subscript $_{(V)}$ to denote that a particular quantity is associated with the effective vielbein $V^a_\mu$. We assume that $V$ can be expanded around $e=f$\footnote{We assume that $V$ is analytic in a neighborhood of the point $e=f$.}
\ba
V(e,f)^a_\mu &=& \sum_{n=0}^\infty \sum_k c^{(V)}_{n,k} (\left[H^n\right]_k)^a_\mu,
\ea
where the $c^{(V)}_{n,k}$ are dimensionless parameters and where $([H^n]_k)^a_\mu$ refers to all possible tensor contractions of $H$ and $e$ at order $n$ in $H$, where $H$ is defined as 
\be
H^a_\mu \equiv e^a_\mu - f^a_\mu.
\ee
We define the covariant derivative $D^{(V)}$, which has the schematic form
\ba
D^{(V)} &\sim& \partial + \Gamma^{(V)} + \om^{(V)}.
\ea
where $\Gamma^{(V),\mu}_{\nu\lambda}$ is the Christoffel connection associated with $V$ (which acts on tangent space indices) and $\om^{(V),ab}_\mu$ is the spin connection (which acts on Lorentz indices). We will see below that $\om^{(V),ab}_\mu$ only enters the calculation for convenience, so we will choose to introduce a connection such that $D^{(V)}_\mu V^a_\nu = 0$.

We will introduce \stu fields following the conventions of \cite{Ondo:2013wka} for later convenience
\ba
e^a_\mu &\rightarrow& \tilde{e}^a_\mu = \Lambda^a_b e^b_\mu \nn \\
f^a_\mu &\rightarrow& \tilde{f}^a_\mu (x)= \partial_\mu \phi^\nu f^a_\nu (\phi(x)).
\ea
where the tilde denotes a quantity that depends on the \stu fields. We define the stress energy tensor with respect to $\tilde{V} \equiv V(\tilde{e},\tilde{f})$ as
\be
\tilde{T}^{(V)}_{\mu\nu}  \equiv  -\frac{2}{|\tilde{V}|} \frac{\delta S^{(V)}_{\rm matt}[\tilde{V},\psi_m]}{\delta \tilde{g}^{(V),\mu\nu}},
\ee
where $|\tilde{V}| \equiv {\rm det}(\tilde{V})$, and where
\be
\tilde{g}^{(V)}_{\mu\nu} \equiv \eta_{ab} \tilde{V}^a_\mu \tilde{V}^b_\nu.
\ee 
The equation of motion for the \stu fields can be written
\be
\frac{\delta S[\tilde{e},\tilde{f},\psi_{\rm m}]}{\delta \phi^a} = m^2 M_{\rm pot}^2 \mathcal{E}^a_{\phi, \rm pot} + \sum_V \mathcal{E}^a_{\phi, (V)} = 0,
\ee 
where $\mathcal{E}^a_{\phi, \rm pot}$ is the usual contribution to the \stu equation of motion from the potential term,
\be
\mathcal{E}^a_{\phi, \rm pot} = |\tilde{f}| \tilde{D}^{(f)}_\mu \left( \tilde{X}_{(f)}^{\mu\nu} \tilde{f}_\nu^a \right),
\ee
where
\be
\label{eq:X-def}
\tilde{X}_{(f)}^{\mu\nu} \equiv - \frac{2}{|\tilde{f}|}\frac{\delta S_{\rm pot}[\tilde{e},\tilde{f}]}{\delta \tilde{g}^{(f)}_{\mu\nu}}.
\ee
To compute $\mathcal{E}^a_{\phi, (V)}$, we vary the matter action with respect to the \stu fields
\ba
\delta S^{(V)}_{{\rm matt}}&=& \int \d^4 x \frac{\delta S^{(V)}_{\rm matt}[\tilde{V},\psi_{\rm m}]}{\delta \tilde{g}^{(V),\mu\nu}}\delta \tilde{g}^{(V),\mu\nu} \nn \\
&=& - \int \d^4 x |\tilde{V}| \tilde{T}^{(V)}_{\mu\nu} \tilde{V}^{\mu}_a \frac{\delta \tilde{V}^{\nu}_a}{\delta \tilde{f}^b_\rho} \tilde{D}^{(f)}_\rho \delta \phi^b ,
\ea 
where we have used the relationship
\be
\label{eq:bigravity-stuckleberg-variation}
\delta \tilde{f}^a_\mu = \tilde{D}^{(f)}_\mu \delta \phi^a = \partial_\mu \delta \phi^a + \tilde{\om}^{(f),a}_{\mu \ \ \  \ b} \delta \phi^b.
\ee
After integrating by parts we can express the contribution of the matter sector associated with $V$ to the \stu equation of motion as
\be
\label{eq:exact-stu-eoms-matter}
\mathcal{E}^a_{\phi,(V)} =- |\tilde{f}| \left( \tilde{D}^{(V)}_\rho \left[ \tilde{T}^{\mu}_{(V), \nu} \tilde{V}^{\nu}_c \frac{\delta \tilde{V}^{c}_\mu}{\delta \tilde{f}^a_\rho} \right] + \left(\tilde{\om}^{(f), a}_{\mu \ \ b} - \tilde{\om}^{(V),a}_{\mu\ \ b}\right) \tilde{T}^{\mu}_{(V),\nu} \tilde{V}^{\nu}_c \frac{\delta \tilde{V}^{c}_\mu}{\delta \tilde{f}^b_\rho} \right).
\ee
So far this is an exact expression. Note that $\tilde{\om}^{(V),ab}_\mu$ actually cancels out of this expression, we include it only so that the diagonalized diff and local Lorentz invariance is manifest.

\subsection{Decoupling limit}

To take the decoupling limit we first perturb around Minkowski
\ba
e^a_\mu &=& \delta^a_\mu + \frac{h^a_\mu}{2 M_e}, \nn \\
f^a_\mu &=& \delta^a_\mu + \frac{k^a_\mu}{2 M_f}, \nn \\
\phi^a &=& x^a + \frac{B^a}{m M_{\rm pot}} + \frac{e^{a\nu} \partial_\nu \pi}{m^2 M_{\rm pot}}, \nn \\
\Lambda^a_{\ \ b} &=& e^{\lambda^a_{\ b} / m M_{\rm pot}} = \delta^a_b + \frac{\lambda^a_{\ \ b}}{m M_{\rm pot}}.
\ea
We then take the decoupling limit by sending
\be
m \rightarrow 0,\ \ M_e, M_f \rightarrow \infty, \ \ {\rm keeping}\ \Lambda_3 \equiv (m^2 M_{\rm pot})^{1/3}\ {\rm fixed.}
\ee
However, this does not fully fix the decoupling limit for the matter couplings. In order to take a decoupling limit, we need to choose a scaling for the parameters $c^{(V)}_{n,k}$ that appear in the effective vielbein. In this section, following \cite{Ondo:2013wka}, we will choose to keep these parameters fixed in the decoupling limit. This is certainly a safe choice, as we are guaranteed that no interactions will be singular in this limit. Furthermore, this gives us insight into the interactions at $\Lambda_3$ of the full theory if these parameters are $O(1)$.

\subsubsection{Ghost for generic non-minimal matter couplings}
With this choice of scaling for $c_{n,k}^{(V)}$, the decoupling limit in the matter sector amounts to the replacement
\be
\label{eq:galileon-expansion}
\tilde{V}^a_\mu \rightarrow V\left(1, 1 +\Pi \right)^a_\mu =\sum_{n,k} (-1)^n c_{n,k}^{(V)} \left([\Pi^n]_k\right)^a_\mu  \ \ {\rm (in\ the\ matter\ sector),}
\ee
where $([\Pi^n]_k)^\mu_\nu$ refers to all possible contractions of $\Pi$ and $\delta$ at order $n$, and where we have defined
\be
\Pi_{\mu\nu} \equiv \frac{\partial_\mu \partial_\nu \pi}{\Lambda_3^3}.
\ee
The Lorentz \stu fields completely decouple from the matter sector in this limit. Thus the metric and vielbein formalisms are equivalent in the decoupling limit, for the matter coupling considered in Equation \eqref{eq:matter-couplings}.
\\

There are several useful simplifying features that occur in the matter sector when we take this decoupling limit. First, since $e^a_\mu = \tilde{e}^a_\mu = \delta^a_\mu$, we can identify local Lorentz and diff indices. Thus in this section we will raise and lower indices with $\eta_{\mu\nu}$. This will require us to be more careful in distinguishing a vielbein from its inverse, thus we will use the notation $[\tilde{V}^{-1}]^\mu_\nu$ to denote the inverse of $\tilde{V}^\mu_\nu$, so
\be
\left[\tilde{V}^{-1}\right]^\mu_\rho \tilde{V}^\rho_\nu = \delta^\mu_\nu.
\ee
Note that $\tilde{V}^\mu_\nu$ will be symmetric, since it is built out of symmetric matrices that commute. Another special feature is that $|\tilde{V}|$ is a scalar, not a tensor density. This is because non-linearly a ratio of densities $|\tilde{V}|/|e|$ is a scalar, but in the decoupling limit limit $|e| = 1$. Finally, we note that many of the connections vanish. In particular, $\Gamma^{(e),\mu}_{\nu \lambda} = \om^{(e),ab}_\mu = \tilde{\om}^{(f),ab}_\mu = 0$. This last condition is very useful. With this the torsion-free condition for $\tilde{f}$ becomes
\be
\tilde{D}^{(f)}_\mu \tilde{f}^a_\nu = \partial_\mu \tilde{f}^a_\nu + \tilde{\Gamma}^{(f),\lambda}_{\mu\nu} \tilde{f}^a_\lambda + \tilde{\om}^{(f),ab}_\mu \tilde{f}^b_\nu =  \partial_\mu \tilde{f}^a_\nu + \tilde{\Gamma}^{(f),\lambda}_{\mu\nu} \tilde{f}^a_\lambda =  0.
\ee
Finally, we note that $\tilde{\Gamma}^{(f),\mu}_{\nu \lambda}$ is a physical quantity in this limit (that is, it need not only appear in covariant derivatives or in curvatures), since a difference in connections $\tilde{\Gamma}^{(f),\mu}_{\nu\lambda} - \Gamma^{(e),\mu}_{\nu\lambda}$ is a tensor, and $\Gamma^{(e),\mu}_{\nu\lambda}=0$.
\\

To compute the equation of motion for the helicity-0 mode $\pi$, we write the variation of the full action as 
\be
\delta S = \int \d^4 x \left( \mathcal{E}_{\phi,{\rm pot}}^a+\sum_V \mathcal{E}^a_{\phi, (V)} \right) e^{\mu}_a \partial_\mu \delta \pi,
\ee
leading to the equation of motion for $\pi$
\be
\mathcal{E}_{\pi} = - \partial_\mu \left( \mathcal{E}^\mu_{\phi,{\rm pot}}+\sum_V \mathcal{E}^\mu_{\phi, (V)} \right) \equiv  \mathcal{E}_{\pi,{\rm pot}} + \sum_V \mathcal{E}_{\pi,(V)}= 0.
\ee
Plugging in the explicit formula for $\mathcal{E}^\mu_{\phi, (V)}$, we find
\be
\label{eq:general-matter-dl}
\mathcal{E}_{\pi, (V)}  = \partial_\alpha \left[ |\tilde{V}|  \partial_\rho \left(\tilde{T}^{\mu}_{(V), \nu} \left[ \tilde{V}^{-1} \right]^{\nu}_\gamma \frac{\delta \tilde{V}^{\gamma}_\mu}{\delta \tilde{f}^\alpha_\rho}\right) + |\tilde{V}| \tilde{\Gamma}^{(V),\rho}_{\rho \lambda}\left(\tilde{T}^{\mu}_{(V), \nu} \left[\tilde{V}^{-1} \right]^{\nu}_\gamma \frac{\delta \tilde{V}^{\gamma}_\mu}{\delta \tilde{f}^\alpha_\lambda}\right)  \right].
\ee
In general Equation \eqref{eq:general-matter-dl} leads to a very complicated equation of motion that will be higher order in derivatives on $\pi$ in the decoupling limit. Note that $\partial \tilde{f} \sim \partial^3 \pi$. Here, there are several sources of derivatives of $\tilde{f}$: fourth derivatives on $\pi$ can arise from second derivatives acting on $\delta \tilde{V} / \delta \tilde{f}$, or from derivatives acting on the connection $\tilde{\Gamma}^{(V)}$. Third derivatives on $\pi$ can arise from derivatives of $\delta \tilde{V} / \delta \tilde{f}$ or from $\tilde{\Gamma}^{(V)}$. Of course, these are precisely the higher derivative terms we would expect from Equation \eqref{eq:pi-matter-interaction}. Furthermore, as discused above, there is no field redefinition that will remove the interactions that give these equations of motion.
\\

As a specific example, let us consider one of the couplings proposed by \cite{Heisenberg:2014rka}
\be
V^a_\mu = \left( \frac{\kappa |e| + |f|}{|f|} \right)^{1/4} f^a_\mu.
\ee
This kind of coupling was analyzed in the decoupling limit in \cite{Heisenberg:2015iqa}, however we do the analysis here because it is instructive to see explicitly how derivatives on $\delta \tilde{V}/ \delta \tilde{f}$ can lead to ghosts. Note that by construction,
\be
|V| = \kappa |e| + |f|.
\ee
In the decoupling limit, $\tilde{V}^a_\mu$ can be written as
\be
\tilde{V}^a_\mu = \Om (\delta^a_\mu + \Pi^a_\mu),
\ee
where
\be
\Om \equiv \left(\frac{\kappa + |1+\Pi|}{|1+\Pi|}\right)^{1/4}.
\ee
The terms that lead to fourth order equations of motion that are universal (in the sense that they do not depend on the specific choice of matter sector) are $\partial^2 \delta V / \delta f$ and $\partial \Gamma$. Explicitly, these are
\ba
\mathcal{E}^{(\partial^4 \pi)}_{\pi,(V)} &=& - |1+ \Pi| \tilde{T}^{(V),\mu}_{\nu} \Big\{ \partial_\alpha \partial_\mu \left[ \Om \left(1 - \frac{\kappa}{4|1+\Pi|} \Om^{-7/4}\right) \left[\tilde{V}^{-1}\right]^{\nu \alpha}   \right] \nn \\
&&\ \ \ \ \ \ \ \ \ \ \  \ \ + \left[\tilde{V}^{-1}\right]^{\nu \alpha}\left[\tilde{V}^{-1}\right]^{\rho\sigma} \Omega \left(1 - \frac{\kappa}{4 |1+\Pi|} \Om^{-7/4} \right) \partial_\alpha \partial_\mu \tilde{V}_{\rho\sigma} \Big\}.
\ea
This will lead to fourth derivatives on $\pi$ that cannot be cancelled for generic matter sectors. This explicitly confirms the argument given in the introduction, that generic non-minimal couplings to matter will create a ghost in the decoupling limit. 

As a check, note that when $\kappa \rightarrow 0$, $V^a_\mu \rightarrow f^a_\mu$ and $\Om \rightarrow 1$. When we set $\kappa = 0$ above, all the fourth derivative terms cancel. This is guaranteed because of the torsion free condition.
\\

We can guarantee that the partial derivatives on the functions $\delta \tilde{V}/\delta \tilde{f}$ vanish by requiring that $V$ is linearly related to $e$ and $f$
\be
\label{eq:linear-V}
V^a_\mu = \alpha_V e^a_\mu + \beta_V f^a_\mu.
\ee
Then the equation of motion in Equation \eqref{eq:general-matter-dl} becomes
\ba
\label{eq:linear-eom}
\mathcal{E}_{\pi,(V)} &=& \beta_V |\tilde{V}| \partial_\rho \left(|\tilde{V}| \tilde{D}^{(V)}_\nu \tilde{T}_{(V)}^{\mu\nu} \tilde{V}_\mu^\rho \right).
\ea
In fact there is still a potential problem, which we can see by focusing on the contribution
\ba
\label{eq:d-of-gamma}
\mathcal{E}_{\pi,(V)} &\supset& \beta_V |\tilde{V}|^2 \tilde{V}_\mu^\rho\  \partial_\rho \left(\tilde{D}^{(V)}_\nu T_{(V)}^{\mu\nu}\right).
\ea
In the last line we have kept all the terms that can potentially have fourth of $\pi$, coming from $\partial \tilde{\Gamma} \sim \partial^4 \pi$. If $\tilde{D}^{(V)}_\nu \tilde{T}^{\mu\nu}$ contains a connection, then this term will generically be non-zero. 
\\

As an example, let us consider the case where we couple a vector field kinetic term to $e$ and a mass term for the vector field to $f$
\be
S_{\rm matt} = \int \d^4 x |e| \left(-\frac{1}{4} g^{\mu\rho}_{(e)} g^{\nu\sigma}_{(e)}F_{\mu\nu} F_{\rho \sigma} \right) + \int \d^4 x |f| \left(- \frac{m_f^2}{2} g_{(f)}^{\mu\nu} A_\mu A_\nu  \right).
\ee
This corresponds to summing over $V = \{e,f\}$, where we choose to take $\alpha_e = \beta_f = 1$ and $\alpha_f = \beta_e = 0$. Then
\be
\tilde{T}_{\mu\nu}^{(f)} = m_f^2 \left( A_\mu A_\nu - \frac{1}{2} \tilde{g}^{(f)}_{\mu\nu} A^2\right).
\ee
where $A^2 \equiv \tilde{g}_{(f)}^{\mu\nu} A_\mu A_\nu$. Note that $\tilde{D}^{(f)}_\mu \tilde{T}^{\mu\nu}_{(f)}$ contains connections
\be
\tilde{D}^{(f)}_\mu \tilde{T}^{\mu\nu}_{(f)} = 2 m_f^2\ \tilde{g}_{(f)}^{\mu \rho}\ \tilde{g}_{(f)}^{\nu \sigma} \left( A_{(\rho} \tilde{D}^{(f)}_{|\mu|} A_{\sigma)} - A_{(\mu} \tilde{D}^{(f)}_{|\sigma|} A_{\rho)}\right).
\ee
A tedious but straightforward calculation shows that we can write the terms of the equation of motion that are fourth order in derivatives on $\pi$ as
\be
\mathcal{E}_{\pi,(V)}^{(\partial^4 \pi)} = - m_f^2 \left[\tilde{f}^{-1} \right]^{\sigma \beta} \left[\tilde{f}^{-1}\right]^{\tau \mu} \tilde{g}_{(f)}^{\rho \omega} A_{\omega} A_{\sigma}\ \partial_\rho \partial_\beta \partial_\mu \partial_\tau \pi.
\ee
These come precisely from the derivatives of the connections that appear in $\tilde{D}^{(f)}_\mu \tilde{T}^{\mu\nu}_{(f)}$.
\\

To avoid this problem, we have several options.
\begin{enumerate}
\item We can minimally couple matter to the vielbein $e$ (so that for all $V$, $\beta_V = 0$).
\item We can require a weak equivalence principle, so that we only couple to a single $V$, in which case the equation of motion for the matter fields will be $D^{(V)}_\mu T^{\mu\nu}_{(V)}=0$. In other words we can minimally couple to $f$, or to a single linear effective vielbein coupling. This is consistent with the result of \cite{Huang:2015yga}.
\item Alternatively, we can put restrictions on the matter field stress-energy tensor $T^{(V)}_{\mu\nu}$ so that the higher derivatives in $\mathcal{E}_{\pi,(V)}$ cancel. 
\end{enumerate}
Let us discuss option 3 in more detail. The simplest way to eliminate the higher derivatives that generically arise in Equation \eqref{eq:d-of-gamma} is to demand that $\tilde{T}^{(V)}_{\mu\nu} = \mathcal{U}^{(V)} \tilde{g}^{(V)}_{\mu\nu}$ where $\mathcal{U}^{(V)}$ is a scalar function of the matter fields. This automatically eliminates the terms that are fourth order in derivatives since there are no connections that appear in $\tilde{D}^{(V)}_{\mu}\tilde{T}_{(V)}^{\mu\nu}$. There are still potentially third derivative terms that could enter from derivatives of $\tilde{V}$
\be
\mathcal{E}_{\pi,(V)} = \beta_V |\tilde{V}| \partial_\mu \left( \partial_\nu \mathcal{U}^{(V)} |\tilde{V}| \left[\tilde{V}^{-1}\right]^{\mu \nu} \right) \supset \beta_V |\tilde{V}| \partial_\nu \mathcal{U}^{(V)} \ \partial_\mu \left( |\tilde{V}| \left[\tilde{V}^{-1} \right]^{\mu\nu} \right).
\ee
However, these potentially dangerous terms vanish, because
\ba
\partial_\mu \left( |\tilde{V}| \left[\tilde{V}^{-1}\right]^{\mu\nu}\right) &=& 2 \beta_V |\tilde{V}| \left( \left[\tilde{V}^{-1}\right]^\mu_{\rho} \left[\tilde{V}^{-1}\right]^\nu_{\sigma} -\left[\tilde{V}^{-1}\right]^\mu_{\sigma} \left[\tilde{V}^{-1}\right]^\nu_{\rho} \right) \partial_\mu \Pi^{\rho\sigma} = 0. \nn \\
\ea
In fact, this can be generalized slightly. The matter equation of motion can generically be written
\be
\sum_V |\tilde{V}| \tilde{D}^{(V)}_\mu \tilde{T}^{\mu\nu}_{(V)} = 0.
\ee
Thus, so long there is only one stress energy tensor, which for definiteness we call $T^{(v)}_{\mu\nu}$, that is not proportional to $g^{(v),\mu\nu}$, then we can use the matter equation of motion to eliminate the contribution of this $T^{(v)}_{\mu\nu}$
\be
|\tilde{v}|\tilde{D}^{(v)}_\mu \tilde{T}^{(v),\mu\nu} = - \sum_{V} |\tilde{V}| \tilde{D}^{(V)}_\mu \left(\mathcal{U}^{(V)} \tilde{g}^{(V),\mu\nu}\right).
\ee
This eliminates the problematic terms coming from $\partial_\nu \tilde{D}_{(v),\mu}(\tilde{T}^{(v),\mu\nu})$, for example a contribution that would arise from a kinetic term.\footnote{This result can also be derived using the Galileon duality, since we can always go to a duality frame where the kinetic term is coupled to $e$ where it is manifestly not coupled to the \stu fields.} Thus, a scalar field with a kinetic term coupled to one effective vielbein, and potentials coupled to any number of effective vielbeins, is ghost-free in the decoupling limit
\be
S = \int \d^4 x |v| \left( -\frac{1}{2} g^{(v),\mu\nu} \partial_\mu \chi \partial_\nu \chi \right) + \sum_V \int \d^4 x  |V| \mathcal{U}_{(V)}(\chi),
\ee
with each $V$ given by a linear effective vielbein as in Equation \eqref{eq:linear-V}. Based on the arguments in the next section, however, the only coupling of this form that will be completely ghost-free in will be the case studied in \cite{Huang:2015yga}, where a the kinetic term is coupled to $e$ or to $f$, and there can be a potential coupled to $e$ and a separate potential coupled to $f$.

There may be other ways to eliminate the higher derivative terms in $\mathcal{E}_{\pi,(V)}$. Nevertheless, we emphasize that these couplings are not generic, and finding a $T^{\mu\nu}_{(V)}$ such that $\mathcal{E}_{\pi,(V)}$ is second order in derivatives is not sufficient to guarantee that this $T^{\mu\nu}_{(V)}$ comes from an action principle. We note that based on Equation \eqref{eq:galileon-expansion}, in the decoupling limit we are considering in this section, a different way to frame the question is how to couple matter to Galileons in a ghost-free way.\footnote{For example, one could consider multi-field extensions of the Galileons \cite{Deffayet:2010zh,Padilla:2010de,Deffayet:2011gz} as inspiration for matter couplings.} 
\\

To summarize, we have found that for a generic coupling to matter, there will be ghosts at $\Lambda_3$. The only known exceptions are either to place special restrictions on the matter coupling, or to couple a generic matter sector to a single, linear effective vielbein.

\section{Cutoff for the linear effective vielbein coupling}
\label{sec:effective-vielbein-dl}
In the previous section, if we did not want to put special restrictions on the matter sector, we were led by a decoupling limit analysis to focus on matter couplings minimally coupled to a single effective vielbein
\be
S_{\rm matt} = S_{\rm matt}[\psi_{\rm m},v].
\ee
where
\be
\label{eq:linear-effective-vielbein}
v^a_\mu = \alpha e^a_\mu + \beta f^a_\mu.
\ee
In fact, as we will show, the interactions of the linear effective vielbein actually vanish in the decoupling limit we have been considering. As a result, in this section we will perform a more detailed analysis tailored to the special linear effective vielbein matter coupling. In what follows we consider a new decoupling limit in which the ghost will appear. This new limit exists (without causing any interactions to go to infinity) because the interactions for the linear effective vielbein coupling vanish in the decoupling limit we have been considering. 

To reiterate, our perspective on the decoupling limit is that it is a scaling limit that we can take in which the theory simplifies. When we take the decoupling limit, we will choose to scale the parameters $\alpha$ and $\beta$ of the matter coupling to force the matter coupling to contribute non-vanishing interactions at $\Lambda_3$. When we do this, we will find a ghost arises in the decoupling limit. After we have found these dangerous operators using the decoupling limit analysis, we can then turn back to the matter coupling away from the decoupling limit (that is, for general $\alpha$ and $\beta$), and place an upper bound on the cutoff of the theory using the ghostly operators we have identified.

\subsection{Vanishing interactions in the original decoupling limit}
An action for a generic matter sector minimally coupled to the linear effective vielbein \eqref{eq:linear-effective-vielbein} can be written
\be
S_{\rm matt} = \int \d^4 x \sqrt{-g^{(v)}} \mathcal{L}[g^{(v)}, \psi],
\ee
where we covariantly couple $\psi$ to $g^{(v)}$. The effective metric is
\be
g^{(v)}_{\mu\nu} = \eta_{ab} v^a_\mu v^b_\nu.
\ee
In the decoupling limit we have been considering, we can write
\be
g^{(v)}_{\mu\nu} = (\alpha+\beta)^2 \left( \eta_{\mu\nu} + \frac{2\beta}{\alpha+\beta} \Pi_{\mu\nu} + \frac{\beta^2}{(\alpha+\beta)^2} \Pi^2_{\mu\nu} \right).
\ee
It is convenient to take $\alpha+\beta=1$. This can be acheived as follows: First rescale the coordinates to get rid of the overall $(\alpha+\beta)^2$ in front of the metric. Then, inside the brackets, we redefine $\beta/(\alpha + \beta) \rightarrow \beta$. Then the effective metric becomes
\be
\label{eq:geff-flat}
g^{(v)}_{\mu\nu} = \eta_{\mu\nu} + 2 \beta \Pi_{\mu\nu} + \beta^2 \Pi^2_{\mu\nu} = \eta_{\alpha \beta} \left(\delta^\alpha_\mu + \beta \Pi^\alpha_\mu\right)\left(\delta^\beta_\nu + \beta \Pi^\beta_\nu \right).
\ee
Now under the Galileon duality transformation \cite{deRham:2013hsa,deRham:2014lqa} 
\ba
x^\mu &\rightarrow& \tilde{x}^\mu = x^\mu - \frac{\beta}{\Lambda_3^3} \partial^\mu \pi,\nn\\
\partial_\mu \pi &\rightarrow& \tilde{\partial}_\mu \rho (\tilde{x}),
\ea
where $\tilde{x}$ and $\rho(\tilde{x})$ are the dual coordinates and dual field respectively, the metric becomes
\be
g^{(v)}_{\mu\nu} \rightarrow \eta_{\mu\nu}.
\ee
This field redefinition removes all interactions between the Galileon and matter, in the limit we are considering. 

\subsection{New decoupling limit analysis}
In spite of this fact, we can still construct a decoupling limit in which the matter interactions do not vanish by choosing a different scaling for $\alpha$ and $\beta$. 
\\

Before taking the decoupling limit, it is worth massaging the expression for the effective metric into a form where the decoupling limit is easy to take. We are not interested in the helicity-2 modes (because as we will see these interactions will arise at a higher scale), so we set
\ba
e^a_\mu &=& \Lambda^a_{\ \ b} \delta^b_\mu \nn \\
f^a_\mu &=& \partial_\mu \left(x^\alpha + \Phi^\alpha \right) \delta^a_\alpha.
\ea
Writing $\Lambda^a_{\ \ b} = \delta^a_b + \lambda^a_{ \ \ b}$ for convenience (we are not perturbing at this stage, we are just reparameterizing $\Lambda$), the effective metric is
\ba
g^{(v)}_{\mu\nu}&=& \eta_{ab} \left( \alpha e^a_\mu + \beta f^a_\mu \right) \left( \alpha e^b_\nu + \beta f^b_\nu \right) \nn \\
&=& (\alpha + \beta)^2 \partial_\mu \left( x^\alpha + \beta \Phi^\alpha \right) \partial_\nu \left( x^\beta + \beta \Phi^\beta \right) \eta_{\alpha\beta} \nn \\
&& \ \ \ \ \ + \alpha \beta \left[ \lambda_{\alpha \mu} \lambda_\nu \Phi^\alpha +  \partial_\mu \Phi^\alpha \lambda_{\alpha \nu} \right],
\ea
where
\be
\lambda_{\mu\nu} \equiv \delta^a_\mu \eta_{ac}\lambda^c_{\ \ d} \delta^d_\nu,
\ee
and where we have used the relationship
\be
\lambda_{\mu\nu} + \lambda_{\nu\mu} + \lambda_{\mu \alpha} \lambda^{\ \alpha}_{\nu} = 0,
\ee
which follows from $\Lambda^a_{\ c}\Lambda^b_{\ d}\eta^{cd} = \eta^{ab}$.
Inspired by the results of the previous section, we perform a diff on the metric
\be
g^{(v)}_{\mu\nu} \rightarrow \tilde{g}_{\mu\nu}^{(v)}(\tilde{x}) = \partial_\mu Y^\alpha  \partial_\nu Y^\beta g^{(v)}_{\alpha \beta}(x(Y)),
\ee
where
\be
Y^\mu \equiv x^\mu + \beta \Phi^\mu.
\ee
After doing this, the effective metric becomes
\be
g^{(v)}_{\mu\nu} = (\alpha + \beta)^2 \eta_{\mu\nu} + \alpha \beta \left[  \lambda_{\alpha \mu} \partial_\nu \Phi^\alpha +  \partial_\mu \Phi^\alpha \lambda_{\alpha \nu} \right],
\ee
where now all the fields and derivatives are evaluated in the new coordinate system. The last term with the Lorentz \stu fields is an obstruction to removing the matter interactions with a coordinate transformation.\footnote{If we had carried the analogous argument out in the metric language, we would find similar non-vanishing terms, ultimately due to the fact that $\partial_\mu B_\nu$ is not a symmetric matrix.} Note that this term is proportional to $\alpha \beta$, so that it is not present if matter is coupled covariantly to a single vielbein.
\\

We will find it convenient now to take $\alpha+\beta=1$ so that the matter is canonically normalized. To take the decoupling limit, we use the scalings 
\ba
\beta &=& \hat{\beta} \sqrt{\frac{\Lambda_3}{m}} \nn \\
\Lambda^a_{\ \ b} &=& e^{\lambda^a_{\ \ b}/(m M_{\rm pot})} = \delta^a_b + \frac{\lambda^a_{\ \ b}}{m M_{\rm pot}} + \cdots \nn \\
\Phi^\alpha &=& \frac{B^\alpha}{m M_{\rm pot}} + \frac{\partial^\alpha \pi}{m^2 M_{\rm pot}},
\ea
where $\lambda^a_{\ \ b}$ is antisymmetric. Taking the limit $m \rightarrow 0, M_{\rm pot}\rightarrow \infty$ holding $\Lambda_3 \equiv (m^2 M_{\rm pot})^{1/3}$ fixed, we find (working in units where $\Lambda_3 = 1$)
\be
\label{eq:geff-dl}
g^{(v)}_{\mu\nu} = \eta_{\mu\nu} + \hat {\beta} (1-\hat{\beta}) \left[  \lambda_{\alpha \mu} \partial^\alpha \partial_\nu \pi +  \partial_\mu \partial^\alpha \pi \lambda_{\alpha \nu} \right].
\ee
The Lorentz \stu field $\lambda$ should be determined by its equation of motion. The exact equation for the Lorentz \stu fields is given by \cite{deRham:2015rxa}
\be
\eta_{ac} f^c_\mu \frac{\partial \mathcal{L}}{\partial f^b_\mu} - \eta_{bc} f^c_\mu \frac{\partial \mathcal{L}}{\partial f^a_\mu} = 0,
\ee
where $\mathcal{L}=\mathcal{L}_{EH} + \mathcal{L}_{mass} + \mathcal{L}_{\rm matter}$. For the effective vielbein coupling, this reduces to
\be
\left[m^2 M_{\rm pot}^2 \mathcal{Q}(e,f) g^{(e),\mu\nu}+ \alpha \beta |v| T^{(v),\mu\nu} \right] \left(e_\mu^a f_\nu^b - e_\mu^b f_\nu^a \right)  = 0,
\ee
where $\mathcal{Q}(e,f)$ is a scalar function that depends on the parameters in the potential term, whose precise form does not matter for our analysis (see \cite{Ondo:2013wka} for a more detailed discussion of the decoupling limit of the potential term). Taking the decoupling limit of this equation amounts to keeping the terms that are $O(1/m)$ on the left hand side. This yields
\be
\label{eq:lorentz-stuck-eom-mc-dl}
\mathcal{Q}(1,1+\Pi) \left[2 \lambda_{\mu\nu} - F_{\mu\nu} - \lambda_{(\mu |\alpha|} \partial^\alpha \partial_{\nu)}  \right] - \hat{\beta}(1-\hat{\beta})\left( - T_{(\mu |\alpha|} \partial^\alpha \partial_{\nu)} \pi \right) = 0,
\ee
where $F_{\mu\nu} \equiv \partial_\mu B_\nu - \partial_\nu B_\mu$. We can solve this perturbatively for $\lambda$
\be
\lambda_{\mu\nu} = \frac{1}{2} F_{\mu\nu} +  O(\varphi^2),
\ee
where $\varphi$ is a generic name for $B$, $\pi$, and the matter fields.

Up to $O(\varphi^2)$, we find that $g^{\rm eff}$ becomes
\ba
\label{eq:geff-dl-mc-order-2}
g^{(v)}_{\mu\nu} = \eta_{\mu\nu} + \frac{ \hat{\beta} (1 - \hat{\beta}) }{2} F_{(\mu|\alpha|} \partial^\alpha \partial_{\nu)}\pi + O(\varphi^3),
\ea
In this form, we see that $g^{(v)}$ is not simply a diff of flat space. In fact, the Riemann tensor associated with this effective metric does not vanish, because of the contributions of the last term. This correponds to the existence of the ghostly operator that arises in this decoupling limit (restoring the dimensions)
\be
S_{\rm d.l.} \supset \int \d^4 x \frac{\hat{\beta}(1-\hat{\beta})}{2 \Lambda_3^5} F_{(\mu|\alpha|} \partial^\alpha \partial_{\nu)}\pi T_{(\eta)}^{\mu\nu},
\ee
where $T^{\mu\nu}_{(\eta)}$ is the Minkowski stress energy tensor of the matter. Thus the equation of motion for the vectors given in Equation \eqref{eq:exact-stu-eoms-matter} will have higher derivatives arising from the connection associated with this metric, that cannot be removed with a field redefiniton. 
\\

This scalar-vector ghost is different from the normal ghosts considered in massive gravity, which involve purely scalar modes. Related to this, we note that it is normally assumed that we can ignore the vectors for solar system tests, because they are not sourced by matter. With this matter coupling, the vectors will sourced by $T^{\mu\nu}_{(\eta)}$ if there is a background for $\pi$.
\\

Notice that to $O(\varphi^2)$ in the decoupling limit, there is no difference between the metric and vielbein formalisms. In other words, to this order, the modification of the symmetric vielbein condition in Equation \eqref{eq:lorentz-stuck-eom-mc-dl} does not enter into the effective metric. Indeed, one can derive Equation \eqref{eq:geff-dl-mc-order-2} by using the metric language. There will be differences between the vielbein and metric languages at higher order, but it is unnecessary for our purposes to calculate these corrections since the ghost already appears at $O(\varphi^2)$. Of course, considering non-minimal covariant couplings (say between matter and $R[g^{(v)}]$) will only make these problems worse. 

\subsection{Upper bound for cutoff scale}
In the previous section we have considered a special, simplifying limit of the theory where the ghost arose immediately at the scale $\Lambda_3$. This limit is a good approximation to the full dynamics of the theory when $\alpha,\beta \sim \sqrt{\Lambda_3/m} \sim (M_{\rm pot}/m)^{1/3} \gg 1$. As a result, we can rule out the possibility of taking $\alpha,\beta$ to be this large based on the analysis of the previous section. 

Of course it is not necessary to take $\alpha,\beta$ to be this large. Nevertheless, we certainly expect that the problems that appear in this limit will still be present even for more general values of $\alpha$ and $\beta$. To make this more concrete, let us consider coupling to a massless scalar field\footnote{Though none of the conclusions we make about the cutoff scale depend on us choosing a massless scalar field.}
\be
S =\int \d^4 x \sqrt{-g^{(v)}} \left(  -\frac{1}{2} g^{(v),\mu\nu} \partial_\mu \chi \partial_\nu \chi\right).
\ee
Without taking a decoupling limit, let's consider the different kinds of interactions that appear when we perturb this around a Minkowski background. We will consider the theory in the metric language for simplicity. Up to quartic order, the action is schematically given by (after scaling coordinates so that $\alpha + \beta =1$)
\ba
S &\sim& \int \d^4 x (\partial \chi)^2 + \left(\frac{h}{\alpha M_e} + \frac{k}{\beta M_f} \right) (\partial \chi)^2 + \left( \frac{h}{\alpha M_e} + \frac{k}{\beta M_f}\right)^2 (\partial \chi)^2 \nn \\
&&+ \frac{\alpha \beta}{m^3 M_{\rm pot}^2} \partial B \partial \partial \pi (\partial \chi)^2 + \frac{\alpha \beta}{m^2 M_{\rm pot}^2} (\partial B)^2 (\partial \chi)^2  + O( \varphi^5),
\ea
where $h_{\mu\nu}$ and $k_{\mu\nu}$ are the fluctuations in $g_{\mu\nu}$ and $f_{\mu\nu}$ respectively. We can see that if $\alpha,\beta \sim (M_{\rm pot}/m)^{1/3}$ then the decoupling limit from the previous section is a good approximation and the ghost will apear at $\Lambda_3$. However, even for general $\alpha,\beta$, from this quartic action we see that scattering processes such as $\chi \chi \rightarrow B \pi$ are dominated by the interaction that we found in the decoupling limit
\be
\label{eq:ghostly-operator}
\mathcal{O}_{(\partial B)(\partial \partial \pi) (\partial \chi)^2} = \frac{\alpha \beta}{m^3 M_{\rm pot}^2} F_{\alpha (\mu} \partial^\alpha \partial_{\nu)} \pi \left( \partial^\mu \chi \partial^\nu \chi -\frac{1}{2} \eta^{\mu\nu} (\partial \chi)^2 \right).
\ee
Because of these interactions are ghostly, we conclude unitarity is broken at or below the scale
\be
\Lambda_{\rm ghost} \sim \frac{(m^3 M_{\rm pot}^2)^{1/5}}{(\alpha \beta)^{1/5}}.
\ee
As a result, we can place a bound $\Lambda_{\rm c.o.} \lesssim \Lambda_{\rm ghost}$. Since this operator also appeared in the vielbein formalism, this bound applies to the matter coupling in the vielbein formalism as well. Taking $m \sim H_0$ (where $H_0\sim 10^{-33}\ {\rm eV}$ is the Hubble parameter), and taking $M_{\rm pot} \sim \mpl$, we find
\be
\Lambda_{\rm c.o.} \lesssim (\alpha \beta)^{-1/5} \times 10^4 \Lambda_3 \sim (\alpha \beta)^{-1/5} \times 10^{-9}\ {\rm eV} \sim (\alpha \beta)^{-1/5} (100\ {\rm m})^{-1}.
\ee
It seems likely that this bound can be improved by considering operators at higher order, however exploring this question is beyond the scope of this work. To discuss physics at distances shorter than $\Lambda_{\rm c.o.}^{-1}$ in this class of theories, a UV completion is needed. In other words, $\Lambda_{\rm c.o.}$ is a genuine cutoff, it is just not a strong coupling scale.
\\

The operator we have found is closely related to the perturbative ADM analysis of \cite{deRham:2014naa}. In that work, integrating out the shift to sixth order in perturbations the Hamiltonian was found to be non-linear in the lapse
\be
\mathcal{H} \supset \frac{(\alpha \beta)^2}{m^2 \mpl^2} (\partial_i \chi)^2 p_\chi^2 N^2.
\ee
It was suggested that we could identify $N \sim \ddot \pi /(\mpl m^2)$ above. If we make this identification, then we would expect an operator of the schematic form
\be
\label{eq:dim-14-op}
\mathcal{O} \sim \left(\frac{\alpha \beta}{m^3 \mpl^2}\right)^2 (\partial_i \chi)^2 p_\chi^2 \ddot \pi^2.
\ee
We see that this operator arises at the same scale that we found in Equation \eqref{eq:ghostly-operator}. It is interesting that we have recovered the same scale by different means. 

In fact, the operator we have found using a fully four dimensional analysis reduces to a covariantized version of Eq. \eqref{eq:dim-14-op} in two spacetime dimensions, in full agreement with the perturbative ADM analysis done in \cite{deRham:2014naa} in two dimensions. To see this, note that in two dimensions we can write
\be
F_{\mu\nu} = f \ep_{\mu\nu},
\ee
for some scalar $f$. Then the important parts of the action are
\be
S \supset \int \d^2 x \left(-\frac{1}{4} F_{\mu\nu} F^{\mu\nu} - \frac{\alpha \beta}{m^3 \mpl^2} (\Pi_{\mu\alpha} F^{\alpha}_{\ \nu} T^{\mu\nu})\right) =\int \d^2 x \left(-\frac{1}{2} f^2 - \frac{\alpha \beta}{m^3 \mpl^2}f (\Pi_{\mu\alpha} \epsilon^{\alpha}_{\ \nu} T^{\mu\nu})\right).
\ee
Integrating out $f$ we find
\be
S \supset \int \d^2 x \left(\frac{\alpha \beta}{m^3 \mpl^2}\right)^2 (\Pi_{\mu\alpha} \epsilon^{\alpha \beta}T^\mu_{\ \beta})^2,
\ee
which is a covariantized version of Eq. \eqref{eq:dim-14-op}. 
\\

Even though unitarity is broken at the scale $\Lambda_{\rm ghost}$, it is still possible to view this matter coupling perturbatively below the cutoff scale. Since our (conservative) upper bound is parametrically larger than $\Lambda_3$ there is still potentially a regime of energies $E$ where $\Lambda_3 < E < \Lambda_{\rm c.o.}$ where the Vainshtein mechanism could potentially operate but we would still be within the regime of the theory. In addition, the fact that a ghost does not appear in the mini-superspace means that we can potentially trust this background to be within the regime of validity of the theory. 

However, we expect a ghost to receive a kinetic term on backgrounds that have anisotropy (and so can excite the above operator at quadratic order in perturbations around that background), including spherically symmetric backgrounds needed for solar system tests. 

Furthermore, if we are willing to view this coupling in a purely perturbative effective field theory sense (with no strong coupling regime), we should include all operators consistent with the symmetries present, not just this matter coupling. For example, while the matter sector is covariant with respect to the effective metric $g^{(v)}$ at tree level, this symmetry is broken by the mass and kinetic terms for $e$ and $f$. Therefore beyond one loop one would expect to generate operators that broke the covariant structure of the matter coupling.
 
Said differently, we cannot trust the interactions from the matter coupling involving the vector modes in a strongly coupled regime. All of the (non-vanishing) vector-scalar-matter interactions will give rise to higher order equations of motion and are ghostly. Thus there is no possibility of a strongly coupled regime for these interactions, a UV completion must be provided at this scale that removes the ghost, and physics above this scale will therefore be sensitive to the UV completion.
\\

To summarize the results of the previous two sections, we have shown that for a very wide class of matter interactions, a ghost arises in the decoupling limit where we scale $\alpha,\beta\sim 1$, in complete agreement with the recent results of \cite{Huang:2015yga,Heisenberg:2015iqa}. For the effective vielbein coupling proposed in \cite{deRham:2014naa}, we have found that in fact, the interactions between $\pi$ and matter vanish in this limit, which we have shown by doing a Galileon duality transformation. Nevertheless, there is still a ghost for these interactions that arises as soon as they interactions become non-trivial. This can be seen by considering a different decoupling limit where we scale $\alpha,\beta \sim 1/\sqrt{m}$ with $\alpha+\beta$ fixed. We can therefore place an upper bound on the cutoff for these theories around a Minkowski background.

\section{Cutoff for new kinetic terms}
\label{sec:kinetic-dl}
With the results of the previous section, we can now quickly consider the possibility of new kinetic interactions in second-order form in the vielbein language. We will work in the massive gravity limit (Equation \eqref{eq:mg-limit}) to simplify the presentation, but the same conclusions when working in bi-gravity.
\subsection{Jordan frame of the linear effective vielbein coupling}
As emphasized in the introduction, we can interpret the new matter coupling as a new kinetic interaction. The Jordan frame of the linear effective vielbein matter coupling is
\be
S = M_{e}^2 S_{EH}[v(e,f_{\rm ref})] + m^2 M_{\rm pot}^2 S_{\rm pot}[e,f_{\rm ref}] + S_{\rm matt}[e,\psi_{\rm m}],
\ee
where the linear effective vielbein is
\be
v^a_\mu = \gamma e^a_\mu + \kappa (f_{\rm ref})^a_\mu.
\ee
The parameters $\gamma,\kappa$ are related to the parameters $\alpha,\beta$ given in the previous section by $\gamma \equiv 1/\alpha, \kappa \equiv \beta/\alpha$. The parameter $\gamma$ changes the canonical normalization for $h$ (the fluctuation of $e$), so it useful to redefine $M_{\rm pot} \equiv \gamma M_e $.

We will perturb the vielbeins in the sense\footnote{Note that we need to keep track of the flucation in $e$ when working with the kinetic term. We will work with the vielbein fluctuation $h^a_\mu$, which is related to the metric perturbation by $h_{\rm vielbein} + \frac{1}{4} h_{\rm vielbein}^2 = h_{\rm metric}$.}
\ba
e^a_\mu &=& (\delta^a_b + \lambda^a_{\ \ b}) \left(\delta^b_\mu + \frac{1}{2} h^a_\mu \right) \nn \\
(f_{\rm ref})^a_\mu &=& \partial_\mu \left(x^\alpha + \Phi^\alpha \right) \delta^a_\alpha.
\ea
Note that we can choose $\lambda^a_{\ b}$ so that $h_{\mu\nu} \equiv \eta_{a\nu} h^a_\mu$ is symmetric. This yields an effective metric
\ba
g^{(v)}_{\mu\nu}&=& \partial_\mu \left( x^\alpha + \kappa \Phi^\alpha \right) \partial_\nu \left( x^\beta + \kappa \Phi^\beta \right) \eta_{\alpha\beta} \nn \\
&& \ \ \ \ \ + \gamma h_{\mu\nu}  + \frac{\gamma^2}{4} h^2_{\mu\nu} +\kappa \gamma \left[ \epsilon_{\alpha \mu} \partial_\nu \Phi^\alpha +  \partial_\mu \Phi^\alpha \epsilon_{\alpha \nu} \right],
\ea
where
\be
\epsilon_{\mu\nu} \equiv \lambda_{\mu\nu} + \frac{1}{2} h_{\mu\nu} +  \lambda_{(\mu |\beta|} h^\beta_{\ \nu)}.
\ee
The kinetic term can be expanded in powers of $h$
\be
M_e^2 S_{EH}[v] =M_e^2 S_{EH}[v]|_{h=0} + M_e^2 \int \d^4 x\ h_{\mu\nu} \left(\mathcal{G}^{(v),\mu\nu}|_{h=0}\right) + O(h^2),
\ee
where $\mathcal{G}^{(v)}_{\mu\nu} \equiv R^{(v)}_{\mu\nu} -\frac{1}{2} g^{(v)}_{\mu\nu}R^{(v)}$
Both of these terms are non-zero because of the piece proportional to $\kappa \gamma$ in the effective metric. Note that in the previous section, the pure scalar interactions vanish in the kinetic term. In this picture, we do not need a Galileon duality transformation to see this, the argument is simply that the effective metric given has vanishing curvature when we only include the scalars, as can be seen directly from Equation \eqref{eq:geff-flat}. These two contributions have the schematic form
\be
M_e^2 S_{EH}[v] \sim \int \d^4 x   (\kappa \gamma)^2\frac{M_e^2 m^2}{\Lambda_3^3}\partial^2 \left(\lambda \partial \partial \pi \right)^2  + (\kappa \gamma) \frac{M_e^2 m^3}{\Lambda_3^6} h \partial^2 (\lambda \partial\partial \pi ).
\ee
The Lorentz \stu equation of motion is
\be
\left[m^2 M_{\rm pot}^2 \mathcal{Q}(e,f) g^{(e),\mu\nu}+ \kappa M_e^2 |v| \mathcal{G}^{(v),\mu\nu} \right] \left(e_\mu^a f_\nu^b - e_\mu^b f_\nu^a \right)  = 0,
\ee
As before, when we integrate out the Lorentz \stu fields, the metric and vielbein formulations are equivalent to $O(\varphi^2)$. Thus, we find a cubic interaction in the decoupling limit arising from the second term above
\be
\mathcal{O}_{(\partial \partial \pi)(\partial A)h} =  \frac{\kappa \gamma}{m \Lambda_3^6}F_{\mu \alpha} \partial^\alpha\partial_\nu \pi (\mathcal{E} h)^{\mu\nu},
\ee
where $\mathcal{E}$ is the Lichnerowicz operator. This will lead to higher order equations of motion. 
\\

While we have derived this operator directly in the language of the kinetic term, it also can be derived simply by starting from the Einstein frame picture and doing field redefinition on $h$ of the form $h \rightarrow h - \partial B \partial \partial \pi$. We can also compute a cutoff scale directly in Jordan frame. We consider the same scattering process as before, $B \pi \rightarrow \chi \chi$. The contribution of the above operator to this process (which also includes the matter coupling, $h_{\mu\nu} T^{\mu\nu} / M_{\rm pot}$) breaks unitarity at the same scale we found before. Of course this is not surprising, as S-matrix elements are invariant under field redefintions.

In fact, this kinetic term was already shown to contain a ghost in the metric language in the decoupling limit analysis of \cite{deRham:2013tfa}. While the details of the decoupling limit are slightly different, nevertheless one can check explicitly using the formalism of that paper that the pure scalar interactions vanish, but the vector-scalar interactions do not. Because the metric and the vielbein are equivalent to leading order in $\varphi$, this also rules out the interaction in the vielbein. 

The above analysis can be easily generalized to bi-gravity, however one has to be careful to first diagonalize the kinetic terms for the fluctuations of $e$ and $f$, since there is kinetic mixing with the modified kinetic term.

\subsection{Einstein-Hilbert of the linear effective vielbein}
We can also consider the new interaction proposed by \cite{Noller:2014ioa}, which as shown in Equation \eqref{eq:one-loop-kt} arises at one loop from the effective vielbein coupling. The interaction is
\be
S= M_e^2 S_{EH}[e] + M_v^2 S_{EH}[v] + M_f^2 S_{EH}[f] + m^2 M_{\rm pot}^2 S_{\rm pot}[v] + S_{\rm matt}[e,\psi_{\rm m}],
\ee
The only effect of the new term, to the order we are working, is to change the normalization of the fluctuation for $e$. The pure scalar interactions cancel, again because when we limit ourselves to the scalars then $g^{(v)}_{\mu\nu}$ has the form of a diff of Minkowski space, so its curvature vanishes. This fact makes this kinetic term a promising candidate for a new kinetic term, a generic modification would already contain a ghost in the pure scalar sector.
\\

Thus we need to include the vectors. However, $S_{EH}[e]$ does not contribute to the Lorentz \stu equations of motion at all, and so the equations are the same as the previous section. Again to $O(\varphi^2)$, the effective vielbein is the same whether calculated in the vielbein or metric formulation. In fact, $S_{EH}[e]$ does not contribute any vector-scalar interactions at all (since it is diff invariant), and so this kinetic interaction has the same ghostly operator we found in the previous section.

\subsection{Simultaneously modifying the kinetic term and the matter coupling}
Finally for completness we consider simultaneously modifying the kinetic term and matter coupling with different effective vielbeins
\be
S = M_e^2 S_{EH}[v(e,f)] + M_f^2 S_{EH}[f] + m^2 M_{\rm pot}^2 S_{\rm pot}[e,f] + S_{\rm matt}[u(e,f),\psi_{\rm m}],
\ee
where $v$ and $u$ are two different linear effective vielbeins. By doing a field redefinition on $e$ we can write this in a form where the fundamental variables are, say, $u$ and $f$
\be
S = M_e^2 S_{EH}[v(u,f)] + M_f^2 S_{EH}[f] + m^2 M_{\rm pot}^2 \hat{S}_{\rm pot}[u,f] + S_{\rm matt}[u,\psi_{\rm m}],
\ee
where $v$ is linearly related to $u$ and $f$, the hat denotes the fact that the parameters in the potential term change under this field redefiniton (but it is still of the ghost-free form). Then by the arguments of this section, there is a ghost unless $v(u,f) = u$.

\section{Discussion}
In this work we considered non-standard kinetic terms and matter couplings in massive gravity and bi-gravity, working in the vielbein language in second-order form. The bottom line is that all known extensions of the matter coupling to a generic matter sector and kinetic term contain ghosts. Furthermore, the ghost or ghosts always appear when we take a decoupling limit in which the extension contributes non-trivial interactions. Beyond the decoupling limit, this translates into an upper bound on the cutoff for these extensions.
\\

We gave an ADM argument that the most promising possibilities for these extensions change the constraint structure so that the BD ghost returns. These arguments are complementary to the ones in \cite{deRham:2015cha} and give a different perspective on why the constraints are lost.
\\

Additionally we considered the scale at which ghostly operators arise in this class of extensions. We considered a very general matter coupling, not assuming a weak equivalence principle, and showed that the generic matter coupling will give rise to ghosts at the scale $\Lambda_3$, arising from interactions between the matter fields and the scalar modes of the graviton. There are special choices of coupling that avoid this conclusion, the only known choice that does not put further restrictions on the matter sector is the linear effective vielbein coupling of \cite{deRham:2014naa}. 
\\

Furthermore, we found using a Galileon duality transformation that the effective vielbein coupling contains no purely scalar-matter interactions. By considering a different decoupling limit where the matter coupling gives rise to genuine interactions, we found that a ghost arises in the decoupling limit and involves vector-scalar-matter interactions. Because of the presence of the vector mode, the operator we find represents a qualitiatively different kind of ghost than the one that usually arises in an arbitrary potential term. Related to this, the matter coupling will source the vector modes if there is a background for the scalar, unlike the case in massive gravity.
\\

The decoupling limit analysis we performed is strictly only a good approximation to the theory if the parameters in the matter coupling are of order $(M_{\rm pot}/m)^{1/3}$. Nevertheless, the ghostly operators we identified in this analysis will still be present in the full theory even beyond this regime. This allowed us to put an upper bound on the scale at which the theory breaks full unitarity and requires a UV completion. We expect that this bound can be improved by considering higher order operators.
\\

This analysis strongly suggests that the linear effective vielbein matter coupling cannot have a strong coupling regime. This is because the equations of motion of the vector-scalar-matter interactions break unitarity: the equations of motion are higher order, and these interactions cannot be removed with a field redefinition since they give rise to a non-zero scattering amplitude. As a result, we can at best consider the matter coupling or kinetic interactions to quadratic order around certain special backgrounds, however working to higher order in perturbations we will inevitably be led to interactions that are beyond the regime of validity of the field theory. Thus, while the matter coupling could be used to describe perturbations around an exact FRW solution, it cannot be trusted to describe structure formation processes that involve going to higher order in perturbations.
\\

This analysis can potentially explain some of the results found previously. It seems likely that the fact that anisotropies are required to see the ghost is related to the fact that the ghost arises from vector-scalar interactions. However, the precise nature of the relationship between these statements is subtle, because the decoupling limit around Minkowski space of \cite{Noller:2014ioa} is the same, yet that theory has a ghost in the mini-superspace. It could be interesting to explore this connection further.
\\

These extensions can still be considered perturbatively below $\Lambda_{\rm c.o.}$. Furthermore, in the case of the vielbein matter copuling, the ghost does not arise perturbatively around the mini-superspace.
\\

This result, combined with other recent results \cite{deRham:2013tfa,deRham:2015rxa,deRham:2015cha,Huang:2015yga,Heisenberg:2015iqa}, places strong constraints on the allowed form of ghost-free massive gravity, bi-gravity, and multi-gravity. It is remarkable that we do not need to invoke diffeomorphism invariance to conclude that we need the Einstein-Hilbert kinetic term and a covariant coupling to matter to avoid the presence of ghost degrees of freedom.

\section*{Acknowledgments}
AAM is supported by an NSF GRFP. I would like to thank Kurt Hinterbichler, Nick Ondo, Johannes Noller, Rachel Rosen, Andrew Tolley for useful discussions, Raquel Ribeiro for useful discussions and comments on the manuscript, and especially Claudia de Rham for many useful discussions, comments on the manuscript, and encouragement.

\appendix

\section*{Appendix}

\section{Review of metric language no-go}
\label{appendix}

In \cite{deRham:2013tfa}, a no-go theorem was given for new kinetic terms in four spacetime dimensions for massive gravity (with the reference metric fixed as $f_{\mu\nu} = \eta_{\mu\nu}$. The interactions considered were of the form
\be
\mathcal{L}_{\rm der}(g, f, \partial g, \partial^2 g),
\ee
with two total derivatives.

Let us take a slightly different approach to explaining the analysis performed in that paper. We can construct a covariant form for the derivative of $g$ by noting
\be
\partial_\lambda g_{\mu\nu} = \frac{1}{2} \left[ g_{\mu\sigma} \Delta^{\sigma}_{\lambda \nu} + (\mu \leftrightarrow \nu) \right],
\ee
where $\Delta$ is the difference in connections
\be
\Delta^\lambda_{\mu\nu} \equiv \Gamma[g]^\lambda_{\mu\nu} - \Gamma[f]^\lambda_{\mu\nu}
\ee
Then the most general action is
\be
\label{eq:kinetic-terms}
S_{\rm der} = \Lambda_{\rm der}^2 \int \d^4 x \sqrt{-g} \mathcal{L}_{\rm der}(g,f,\Delta,R),
\ee
where again, each term in $\mathcal{L}_{\rm der}$ must have a total of two derivatives. The scale $\Lambda_{\rm der}^2$ is in principle arbitrary. We want to study the leading order interactions, so we will scale $\Lambda_{\rm der}$ in such a way that the leading order interactions survive in the $\Lambda_3$ decoupling limit\footnote{This choice of scaling for $\Lambda_{\rm der}$ is just for convenience. The real point is to study the leading order interactions.}.

We then choose to introduce \stu fields. We can introduce them through $g$ as
\be
g_{\mu\nu} \rightarrow \tilde{g}_{\mu\nu} = \partial_\mu \phi^\alpha \partial_\nu \phi^\beta g_{\alpha\beta}(\phi(x)).
\ee
Then defining
\be
H_{\mu\nu} \equiv g_{\mu\nu} - \eta_{\mu\nu},
\ee
we can express the action as
\be
\label{eq:Lder}
\mathcal{L}_{\rm der} \sim \sum_n \partial^2 H^n,
\ee
with all indices contracted with the flat reference metric $\eta_{\mu\nu}.$ Note that, unlike the analysis of \cite{deRham:2010ik}, when the \stu fields are introduced through $g$ the helicity-2 mode $h_{\mu\nu}$ only ever appears in $H_{\mu\nu}$.
Then perturbing $\phi^\alpha = x^\alpha + (m \mpl)^{-1} B^\alpha + (m^2 \mpl)^{-1}\partial^\alpha \pi$, we can write $H_{\mu\nu}$ as\footnote{Note that there were typos in early versions of \cite{deRham:2013tfa}, this is the correct version of the \stu prescription that was used in the calculation.}
\ba
H_{\mu\nu}& = & \frac{h_{\mu\nu}}{\mpl} + \frac{ \partial_{(\mu} B_{\nu)} }{m \mpl}+ 2 \frac{ \partial_\mu \partial_\nu \pi}{m^2 \mpl}  \\
&& + \left(\frac{\partial_\mu B^\alpha}{m \mpl} + \frac{\partial_\mu \partial^\alpha \pi}{m^2 \mpl}\right)\left( \frac{\partial_\nu B_\alpha}{m \mpl} + \frac{\partial_\nu \partial_\alpha \pi}{m^2 \mpl} \right), \nn
\ea
with indices raised and lowered with $\eta_{\mu\nu}$. We have neglected the fact that $h_{\mu\nu}$ depends on $\phi$ (which corresponds to replacing $h_{\mu\nu}\rightarrow \partial_\mu \phi^\alpha \partial_\nu \phi^\beta h_{\alpha \beta}(\phi)$ above). In the decoupling limit that we consider, this can be shown to be irrelevant, for more details see \cite{deRham:2013tfa}. \\

We then demand that the equation of motion of $h_{\mu\nu}$, $B_\mu$, and $\pi$ be second order. This is referred to as Property 2 in \cite{deRham:2013tfa}.
Applied to the derivative interactions, this is an extremely restrictive condition. It demands that the helicity-0 mode $\pi$ not enter into the decoupling limit at all, since any term in the action with $\pi$ comes with at least 4 derivatives per field ($\partial^2  h \partial^2 \pi$), so the equations of motion are automatically higher order if they are non-vanishing. 
\begin{itemize}
\item First note that the Einstein Hilbert term gives rise to a kinetic term
\be
\mathcal{L}_{\rm FP} \sim h \partial^2 h.
\ee
To be ghost-free, this must take the form of the Fierz-Pauli kinetic term.

\item The first interactions that can arise in the kinetc term are purely $\pi$ interactions. We can focus on these interactions by choosing $\Lambda_{\rm der} \sim \Lambda_3$,
\be
\mathcal{L}_{{\rm der}, (m^2\mpl)^{1/3}} \sim \Lambda_3^2\  \partial^2 \left(\frac{\partial^2 \pi}{\Lambda_3^3} \right)^{n_\pi}.
\ee
These interactions automatically give higher order equations of motion if they are not total derivatives. Furthermore there is no hope of mixing equations of motion at this scale. The coefficients must be chosen to remove these interactions entirely.

\item Once we cancel these interactions, we choose $\Lambda_{\rm der} \sim (m \mpl^2)^{1/3}$, yielding
\be
\mathcal{L}_{{\rm der}, (m \mpl^2)^{1/3}} \sim \Lambda_3  h\  \partial^2 \left(\frac{\partial^2 \pi}{\Lambda_3^3}\right)^n, \ \ (\partial B)^2\ \partial^2 \left(\frac{\partial^2 \pi}{\Lambda_3^2}\right)^{n_\pi}.
\ee
These interactions again must be cancelled, which fixes more coefficients.

\item The next interactions arise at the scale $\Lambda_{\rm der}\sim (m \mpl^5)^{1/6}$. These have the form
\be
\mathcal{L}_{{\rm der}, (m \mpl^5)^{1/6}} \sim \frac{1}{\Lambda_3^5}\partial^2 h \partial^2 (\partial B)^2, \ \ \frac{1}{\Lambda_3^4}\ \partial^2 (\partial B)^3
\ee
This scaling was not considered in \cite{deRham:2013tfa}. However, we recently performed an analysis, new for this work, demanding that the equations of motion were second order when scaling $\Lambda_{\rm der} \sim (m \mpl^5)^{1/6}$ (working with the most general lagrangian of the form \eqref{eq:Lder} up to quartic order in $H$). The only kinetic interactions that survive this decoupling limit analysis are (1) the Einstein-Hilbert term, and (2) the interaction $\mathcal{L}_{2,4} \sim \ep \ep \partial^2 H^4$ proposed by \cite{Hinterbichler:2013eza} (which is a total derivative in four dimensions). In order for (2) to be represent a genuinely new kinetic interaction, one would still need to find a ghost-free non-linear completion, which we do not exist since the decoupling limit analysis showed there was no non-linear completion for the closely related $\mathcal{L}_{2,3}$.  However this point is moot in four dimensions since $\mathcal{L}_{2,4}$ is a total derivative. As a result, in four dimensions, the decoupling limit analysis is strong enough to show that the Einstein-Hilbert term is the unique ghost-free decoupling limit up to quartic order in $H$. We will now, however, return to reviewing the analysis of \cite{deRham:2013tfa}.
\end{itemize}

Using this method, the most general action of the form given in Equation \eqref{eq:kinetic-terms} that has second order equations of motion in the decoupling limit up to quartic order in $H$. This four parameter family of interaction includes the Einstein-Hilbert term, as well as a candidate non-linear completion of the term proposed in \cite{Hinterbichler:2013eza}, extended to cubic order. By construction, all of these interactions are trivial in the decoupling limit taking $\Lambda_{\rm der} \sim (m \mpl^2)^{1/3}$, but differ beyond that.
\be
\mathcal{L}^{(4)}_{\rm der,\ g.f.\ in\ d.l.}=a \mathcal{L}_a + b_1 \mathcal{L}_{b_1} + b_2 \mathcal{L}_{b_2} + c_1 \mathcal{L}_{c_1},
\ee
where the Lagrangians are all quartic order in $H$. $\mathcal{L}_a$ starts at quadratic order $\mathcal{L}_a \sim \partial^2 \left(H^2 + H^3 + H^4\right)$. Meanwhile $\mathcal{L}_{b_{1,2}}$ start at cubic order $\sim \partial^2\left(H^3 + H^4 \right)$, and $\mathcal{L}_{c_4}$ is purely quartic order $\sim \partial^2 H^4$. 

The remaining four parameter family was then restricted by a perturbative Hamiltonian analysis. In the mini-superspace approximation, two of the terms appear quadratic in the lapse. The final parameter was eliminated by considering anisotropies, where it was shown that the shift picks up a kinetic term and becomes dynamical.
After a perturbative ADM analysis:
\be
\mathcal{L}^{(4)}_{g.f.} = a  \left( \mathcal{L}_a - \mathcal{L}_{b_1} - 4 \mathcal{L}_{b_2} \right) =  a \mathcal{L}_{EH}.
\ee
which is just the Einstein Hilbert term expanded to quartic order.

It was then argued that there are no interactions at any order in perturbation theory. This follows from the seemingly innocuous statement that at leading order in the helicity decomposition, any non-linear completion must be ghost-free. In other words, the interaction must be ghost-free in the decoupling limit at leading order in perturbation theory, without any possibility of mixing orders.
\\

In \cite{Noller:2014ioa}, it was pointed out that field redefinitions are apparently not covered in the above analysis. For example, we can imagine rescalings of $H$
\be
H_{\mu\nu} \rightarrow \lambda H_{\mu\nu}.
\ee
Then if we introduce \stu fields through $H$, we find
\be
\lambda H_{\mu\nu} = \lambda \frac{ h_{\mu\nu}}{\mpl} + 2 \lambda \Psi_{(\mu\nu)} + \lambda \Psi_{\mu \alpha} \Psi^{\ \ \alpha}_\nu,
\ee
where
\be
\Psi_{\mu\nu} \equiv \frac{\partial_{\mu}B_{\nu}}{m \mpl} + \frac{\partial_\mu \partial_\nu \pi}{m^2 \mpl}.
\ee
Because this decomposition does not preserve non-linear diffs, there will be higher order equations of motion on the \stu fields. Of course, these higher order equations of motion do not lead to ghosts, since all we have done is perform a field redefinition on Einstein-Hilbert. To make this explicit, we can perform a field redefinition on $h$ to remove all higher deritvatives from the equations of motion
\be
\label{eq:h-redef}
h_{\mu\nu} \rightarrow h_{\mu\nu} + \lambda(\lambda-1) \mpl \Psi_{\mu \alpha} \Psi^{\ \ \alpha}_\nu
\ee
then rescaling $h_{\mu\nu}\rightarrow h_{\mu\nu}/\lambda$, $B_\mu \rightarrow B_\mu/\lambda$, and $\pi \rightarrow \pi / \lambda.$

However, there is subtlety if matter is coupled to $g_{\mu\nu}=\eta_{\mu\nu}+H_{\mu\nu}$ in the frame where the kinetic term depends on the metric $g'_{\mu\nu} = \eta_{\mu\nu} + \lambda H_{\mu\nu}$. The same field redefiniton we have done on $h_{\mu\nu}$ to remove the higher derivatives from the kinetic term in Equation \eqref{eq:h-redef} will introduce \stu fields into the matter sector. To analyze the decoupling limit in \cite{deRham:2013tfa}, it was implicitly assumed the metric appearing in the kinetic interactions was the same metric appearing in the matter coupling. Then demanding that no \stu fields appear in the matter coupling fixed the freedom to do field redefintions on $H_{\mu\nu}$. Similarly, we can consider arbitary nonlinear field redefinitions of $H_{\mu\nu}$. This can be absorbed into a (highly nonlinear) redefinition of $h_{\mu\nu}$. As before, this will introduce \stu fields into the matter coupling.

\bibliographystyle{JHEPmodplain}
\bibliography{refs}

\providecommand{\href}[2]{#2}\begingroup\raggedright\begin{thebibliography}{100}

\bibitem{Hinterbichler:2011tt}
K.~Hinterbichler, {\it {Theoretical Aspects of Massive Gravity}},  {\sl
  Rev.Mod.Phys.} {\bf 84} (2012) 671--710,
  [\href{http://arxiv.org/abs/1105.3735}{{\sf arXiv:1105.3735}}],
  [\href{http://dx.doi.org/10.1103/RevModPhys.84.671}{{\sf
  doi:10.1103/RevModPhys.84.671}}].

\bibitem{deRham:2014zqa}
C.~de~Rham, {\it {Massive Gravity}},  {\sl Living Rev.Rel.} {\bf 17} (2014) 7,
  [\href{http://arxiv.org/abs/1401.4173}{{\sf arXiv:1401.4173}}],
  [\href{http://dx.doi.org/10.12942/lrr-2014-7}{{\sf
  doi:10.12942/lrr-2014-7}}].

\bibitem{deRham:2010ik}
C.~de~Rham and G.~Gabadadze, {\it {Generalization of the Fierz-Pauli Action}},
  {\sl Phys.Rev.} {\bf D82} (2010) 044020,
  [\href{http://arxiv.org/abs/1007.0443}{{\sf arXiv:1007.0443}}],
  [\href{http://dx.doi.org/10.1103/PhysRevD.82.044020}{{\sf
  doi:10.1103/PhysRevD.82.044020}}].

\bibitem{deRham:2010kj}
C.~de~Rham, G.~Gabadadze, and A.~J. Tolley, {\it {Resummation of Massive
  Gravity}},  {\sl Phys.Rev.Lett.} {\bf 106} (2011) 231101,
  [\href{http://arxiv.org/abs/1011.1232}{{\sf arXiv:1011.1232}}],
  [\href{http://dx.doi.org/10.1103/PhysRevLett.106.231101}{{\sf
  doi:10.1103/PhysRevLett.106.231101}}].

\bibitem{Hassan:2011zd}
S.~Hassan and R.~A. Rosen, {\it {Bimetric Gravity from Ghost-free Massive
  Gravity}},  {\sl JHEP} {\bf 1202} (2012) 126,
  [\href{http://arxiv.org/abs/1109.3515}{{\sf arXiv:1109.3515}}],
  [\href{http://dx.doi.org/10.1007/JHEP02(2012)126}{{\sf
  doi:10.1007/JHEP02(2012)126}}].

\bibitem{Hinterbichler:2012cn}
K.~Hinterbichler and R.~A. Rosen, {\it {Interacting Spin-2 Fields}},  {\sl
  JHEP} {\bf 1207} (2012) 047, [\href{http://arxiv.org/abs/1203.5783}{{\sf
  arXiv:1203.5783}}], [\href{http://dx.doi.org/10.1007/JHEP07(2012)047}{{\sf
  doi:10.1007/JHEP07(2012)047}}].

\bibitem{Hassan:2012wt}
S.~Hassan, A.~Schmidt-May, and M.~von Strauss, {\it {Metric Formulation of
  Ghost-Free Multivielbein Theory}},
  \href{http://arxiv.org/abs/1204.5202}{{\sf arXiv:1204.5202}}.

\bibitem{Fierz:1939ix}
M.~Fierz and W.~Pauli, {\it {On relativistic wave equations for particles of
  arbitrary spin in an electromagnetic field}},  {\sl Proc.Roy.Soc.Lond.} {\bf
  A173} (1939) 211--232, [\href{http://dx.doi.org/10.1098/rspa.1939.0140}{{\sf
  doi:10.1098/rspa.1939.0140}}].

\bibitem{Boulware:1973my}
D.~Boulware and S.~Deser, {\it {Can gravitation have a finite range?}},  {\sl
  Phys.Rev.} {\bf D6} (1972) 3368--3382,
  [\href{http://dx.doi.org/10.1103/PhysRevD.6.3368}{{\sf
  doi:10.1103/PhysRevD.6.3368}}].

\bibitem{Hassan:2012qv}
S.~Hassan, A.~Schmidt-May, and M.~von Strauss, {\it {Proof of Consistency of
  Nonlinear Massive Gravity in the St\'uckelberg Formulation}},  {\sl
  Phys.Lett.} {\bf B715} (2012) 335--339,
  [\href{http://arxiv.org/abs/1203.5283}{{\sf arXiv:1203.5283}}],
  [\href{http://dx.doi.org/10.1016/j.physletb.2012.07.018}{{\sf
  doi:10.1016/j.physletb.2012.07.018}}].

\bibitem{Mirbabayi:2011aa}
M.~Mirbabayi, {\it {A Proof Of Ghost Freedom In de Rham-Gabadadze-Tolley
  Massive Gravity}},  {\sl Phys.Rev.} {\bf D86} (2012) 084006,
  [\href{http://arxiv.org/abs/1112.1435}{{\sf arXiv:1112.1435}}],
  [\href{http://dx.doi.org/10.1103/PhysRevD.86.084006}{{\sf
  doi:10.1103/PhysRevD.86.084006}}].

\bibitem{Hassan:2011hr}
S.~Hassan and R.~A. Rosen, {\it {Resolving the Ghost Problem in non-Linear
  Massive Gravity}},  {\sl Phys.Rev.Lett.} {\bf 108} (2012) 041101,
  [\href{http://arxiv.org/abs/1106.3344}{{\sf arXiv:1106.3344}}],
  [\href{http://dx.doi.org/10.1103/PhysRevLett.108.041101}{{\sf
  doi:10.1103/PhysRevLett.108.041101}}].

\bibitem{Hassan:2011ea}
S.~Hassan and R.~A. Rosen, {\it {Confirmation of the Secondary Constraint and
  Absence of Ghost in Massive Gravity and Bimetric Gravity}},  {\sl JHEP} {\bf
  1204} (2012) 123, [\href{http://arxiv.org/abs/1111.2070}{{\sf
  arXiv:1111.2070}}], [\href{http://dx.doi.org/10.1007/JHEP04(2012)123}{{\sf
  doi:10.1007/JHEP04(2012)123}}].

\bibitem{Hassan:2011tf}
S.~Hassan, R.~A. Rosen, and A.~Schmidt-May, {\it {Ghost-free Massive Gravity
  with a General Reference Metric}},  {\sl JHEP} {\bf 1202} (2012) 026,
  [\href{http://arxiv.org/abs/1109.3230}{{\sf arXiv:1109.3230}}],
  [\href{http://dx.doi.org/10.1007/JHEP02(2012)026}{{\sf
  doi:10.1007/JHEP02(2012)026}}].

\bibitem{Hassan:2011vm}
S.~Hassan and R.~A. Rosen, {\it {On Non-Linear Actions for Massive Gravity}},
  {\sl JHEP} {\bf 1107} (2011) 009, [\href{http://arxiv.org/abs/1103.6055}{{\sf
  arXiv:1103.6055}}], [\href{http://dx.doi.org/10.1007/JHEP07(2011)009}{{\sf
  doi:10.1007/JHEP07(2011)009}}].

\bibitem{deRham:2011qq}
C.~de~Rham, G.~Gabadadze, and A.~J. Tolley, {\it {Helicity Decomposition of
  Ghost-free Massive Gravity}},  {\sl JHEP} {\bf 1111} (2011) 093,
  [\href{http://arxiv.org/abs/1108.4521}{{\sf arXiv:1108.4521}}],
  [\href{http://dx.doi.org/10.1007/JHEP11(2011)093}{{\sf
  doi:10.1007/JHEP11(2011)093}}].

\bibitem{deRham:2011rn}
C.~de~Rham, G.~Gabadadze, and A.~J. Tolley, {\it {Ghost free Massive Gravity in
  the St\"uckelberg language}},  {\sl Phys.Lett.} {\bf B711} (2012) 190--195,
  [\href{http://arxiv.org/abs/1107.3820}{{\sf arXiv:1107.3820}}],
  [\href{http://dx.doi.org/10.1016/j.physletb.2012.03.081}{{\sf
  doi:10.1016/j.physletb.2012.03.081}}].

\bibitem{Kugo:2014hja}
T.~Kugo and N.~Ohta, {\it {Covariant Approach to the No-ghost Theorem in
  Massive Gravity}},  {\sl PTEP} {\bf 2014} (2014), no.~4 043B04,
  [\href{http://arxiv.org/abs/1401.3873}{{\sf arXiv:1401.3873}}],
  [\href{http://dx.doi.org/10.1093/ptep/ptu046}{{\sf
  doi:10.1093/ptep/ptu046}}].

\bibitem{deRham:2010tw}
C.~de~Rham, G.~Gabadadze, L.~Heisenberg, and D.~Pirtskhalava, {\it {Cosmic
  Acceleration and the Helicity-0 Graviton}},  {\sl Phys.Rev.} {\bf D83} (2011)
  103516, [\href{http://arxiv.org/abs/1010.1780}{{\sf arXiv:1010.1780}}],
  [\href{http://dx.doi.org/10.1103/PhysRevD.83.103516}{{\sf
  doi:10.1103/PhysRevD.83.103516}}].

\bibitem{D'Amico:2011jj}
G.~D'Amico, C.~de~Rham, S.~Dubovsky, G.~Gabadadze, D.~Pirtskhalava, {\em
  et~al.}, {\it {Massive Cosmologies}},  {\sl Phys.Rev.} {\bf D84} (2011)
  124046, [\href{http://arxiv.org/abs/1108.5231}{{\sf arXiv:1108.5231}}],
  [\href{http://dx.doi.org/10.1103/PhysRevD.84.124046}{{\sf
  doi:10.1103/PhysRevD.84.124046}}].

\bibitem{vonStrauss:2011mq}
M.~von Strauss, A.~Schmidt-May, J.~Enander, E.~Mortsell, and S.~Hassan, {\it
  {Cosmological Solutions in Bimetric Gravity and their Observational Tests}},
  {\sl JCAP} {\bf 1203} (2012) 042, [\href{http://arxiv.org/abs/1111.1655}{{\sf
  arXiv:1111.1655}}],
  [\href{http://dx.doi.org/10.1088/1475-7516/2012/03/042}{{\sf
  doi:10.1088/1475-7516/2012/03/042}}].

\bibitem{Volkov:2011an}
M.~S. Volkov, {\it {Cosmological solutions with massive gravitons in the
  bigravity theory}},  {\sl JHEP} {\bf 1201} (2012) 035,
  [\href{http://arxiv.org/abs/1110.6153}{{\sf arXiv:1110.6153}}],
  [\href{http://dx.doi.org/10.1007/JHEP01(2012)035}{{\sf
  doi:10.1007/JHEP01(2012)035}}].

\bibitem{Comelli:2011zm}
D.~Comelli, M.~Crisostomi, F.~Nesti, and L.~Pilo, {\it {FRW Cosmology in Ghost
  Free Massive Gravity}},  {\sl JHEP} {\bf 1203} (2012) 067,
  [\href{http://arxiv.org/abs/1111.1983}{{\sf arXiv:1111.1983}}],
  [\href{http://dx.doi.org/10.1007/JHEP06(2012)020,
  10.1007/JHEP03(2012)067}{{\sf doi:10.1007/JHEP06(2012)020,
  10.1007/JHEP03(2012)067}}].

\bibitem{Chiang:2012vh}
C.-I. Chiang, K.~Izumi, and P.~Chen, {\it {Spherically symmetric analysis on
  open FLRW solution in non-linear massive gravity}},  {\sl JCAP} {\bf 1212}
  (2012) 025, [\href{http://arxiv.org/abs/1208.1222}{{\sf arXiv:1208.1222}}],
  [\href{http://dx.doi.org/10.1088/1475-7516/2012/12/025}{{\sf
  doi:10.1088/1475-7516/2012/12/025}}].

\bibitem{Akrami:2012vf}
Y.~Akrami, T.~S. Koivisto, and M.~Sandstad, {\it {Accelerated expansion from
  ghost-free bigravity: a statistical analysis with improved generality}},
  {\sl JHEP} {\bf 1303} (2013) 099, [\href{http://arxiv.org/abs/1209.0457}{{\sf
  arXiv:1209.0457}}], [\href{http://dx.doi.org/10.1007/JHEP03(2013)099}{{\sf
  doi:10.1007/JHEP03(2013)099}}].

\bibitem{Tasinato:2012ze}
G.~Tasinato, K.~Koyama, and G.~Niz, {\it {Vector instabilities and
  self-acceleration in the decoupling limit of massive gravity}},  {\sl
  Phys.Rev.} {\bf D87} (2013), no.~6 064029,
  [\href{http://arxiv.org/abs/1210.3627}{{\sf arXiv:1210.3627}}],
  [\href{http://dx.doi.org/10.1103/PhysRevD.87.064029}{{\sf
  doi:10.1103/PhysRevD.87.064029}}].

\bibitem{DeFelice:2013bxa}
A.~De~Felice, A.~E. G{\"u}mr{\"u}k\c{c}{\"u}o\u{g}lu, C.~Lin, and S.~Mukohyama,
  {\it {On the cosmology of massive gravity}},  {\sl Class.Quant.Grav.} {\bf
  30} (2013) 184004, [\href{http://arxiv.org/abs/1304.0484}{{\sf
  arXiv:1304.0484}}],
  [\href{http://dx.doi.org/10.1088/0264-9381/30/18/184004}{{\sf
  doi:10.1088/0264-9381/30/18/184004}}].

\bibitem{Akrami:2013pna}
Y.~Akrami, T.~S. Koivisto, and M.~Sandstad, {\it {Cosmological constraints on
  ghost-free bigravity: background dynamics and late-time acceleration}},
  \href{http://arxiv.org/abs/1302.5268}{{\sf arXiv:1302.5268}},
  \href{http://dx.doi.org/10.1142/9789814623995-0138}{{\sf
  doi:10.1142/9789814623995-0138}}.

\bibitem{Volkov:2013roa}
M.~S. Volkov, {\it {Self-accelerating cosmologies and hairy black holes in
  ghost-free bigravity and massive gravity}},  {\sl Class.Quant.Grav.} {\bf 30}
  (2013) 184009, [\href{http://arxiv.org/abs/1304.0238}{{\sf
  arXiv:1304.0238}}],
  [\href{http://dx.doi.org/10.1088/0264-9381/30/18/184009}{{\sf
  doi:10.1088/0264-9381/30/18/184009}}].

\bibitem{Andrews:2013uca}
M.~Andrews, K.~Hinterbichler, J.~Stokes, and M.~Trodden, {\it {Cosmological
  perturbations of massive gravity coupled to DBI Galileons}},  {\sl
  Class.Quant.Grav.} {\bf 30} (2013) 184006,
  [\href{http://arxiv.org/abs/1306.5743}{{\sf arXiv:1306.5743}}],
  [\href{http://dx.doi.org/10.1088/0264-9381/30/18/184006}{{\sf
  doi:10.1088/0264-9381/30/18/184006}}].

\bibitem{Do:2013tea}
T.~Q. Do and W.~Kao, {\it {Anisotropically expanding universe in massive
  gravity}},  {\sl Phys.Rev.} {\bf D88} (2013), no.~6 063006,
  [\href{http://dx.doi.org/10.1103/PhysRevD.88.063006}{{\sf
  doi:10.1103/PhysRevD.88.063006}}].

\bibitem{Bamba:2013hza}
K.~Bamba, Y.~Kokusho, S.~Nojiri, and N.~Shirai, {\it {Cosmology and stability
  in scalar–tensor bigravity}},  {\sl Class.Quant.Grav.} {\bf 31} (2014)
  075016, [\href{http://arxiv.org/abs/1310.1460}{{\sf arXiv:1310.1460}}],
  [\href{http://dx.doi.org/10.1088/0264-9381/31/7/075016}{{\sf
  doi:10.1088/0264-9381/31/7/075016}}].

\bibitem{Konnig:2013gxa}
F.~Koennig, A.~Patil, and L.~Amendola, {\it {Viable cosmological solutions in
  massive bimetric gravity}},  {\sl JCAP} {\bf 1403} (2014) 029,
  [\href{http://arxiv.org/abs/1312.3208}{{\sf arXiv:1312.3208}}],
  [\href{http://dx.doi.org/10.1088/1475-7516/2014/03/029}{{\sf
  doi:10.1088/1475-7516/2014/03/029}}].

\bibitem{Guarato:2013gba}
P.~Guarato and R.~Durrer, {\it {Perturbations for massive gravity theories}},
  {\sl Phys.Rev.} {\bf D89} (2014), no.~8 084016,
  [\href{http://arxiv.org/abs/1309.2245}{{\sf arXiv:1309.2245}}],
  [\href{http://dx.doi.org/10.1103/PhysRevD.89.084016}{{\sf
  doi:10.1103/PhysRevD.89.084016}}].

\bibitem{Konnig:2014dna}
F.~Könnig and L.~Amendola, {\it {Instability in a minimal bimetric gravity
  model}},  {\sl Phys.Rev.} {\bf D90} (2014) 044030,
  [\href{http://arxiv.org/abs/1402.1988}{{\sf arXiv:1402.1988}}],
  [\href{http://dx.doi.org/10.1103/PhysRevD.90.044030}{{\sf
  doi:10.1103/PhysRevD.90.044030}}].

\bibitem{Goon:2014ywa}
G.~Goon, A.~E. G{\"u}mr{\"u}k\c{c}{\"u}o\u{g}lu, K.~Hinterbichler,
  S.~Mukohyama, and M.~Trodden, {\it {Galileons Coupled to Massive Gravity:
  General Analysis and Cosmological Solutions}},  {\sl JCAP} {\bf 1408} (2014)
  008, [\href{http://arxiv.org/abs/1402.5424}{{\sf arXiv:1402.5424}}],
  [\href{http://dx.doi.org/10.1088/1475-7516/2014/08/008}{{\sf
  doi:10.1088/1475-7516/2014/08/008}}].

\bibitem{Comelli:2014bqa}
D.~Comelli, M.~Crisostomi, and L.~Pilo, {\it {FRW Cosmological Perturbations in
  Massive Bigravity}},  {\sl Phys.Rev.} {\bf D90} (2014) 084003,
  [\href{http://arxiv.org/abs/1403.5679}{{\sf arXiv:1403.5679}}],
  [\href{http://dx.doi.org/10.1103/PhysRevD.90.084003}{{\sf
  doi:10.1103/PhysRevD.90.084003}}].

\bibitem{Solomon:2014dua}
A.~R. Solomon, Y.~Akrami, and T.~S. Koivisto, {\it {Linear growth of structure
  in massive bigravity}},  {\sl JCAP} {\bf 1410} (2014) 066,
  [\href{http://arxiv.org/abs/1404.4061}{{\sf arXiv:1404.4061}}],
  [\href{http://dx.doi.org/10.1088/1475-7516/2014/10/066}{{\sf
  doi:10.1088/1475-7516/2014/10/066}}].

\bibitem{DeFelice:2014nja}
A.~De~Felice, A.~E. G{\"u}mr{\"u}k\c{c}{\"u}o\u{g}lu, S.~Mukohyama,
  N.~Tanahashi, and T.~Tanaka, {\it {Viable cosmology in bimetric theory}},
  {\sl JCAP} {\bf 1406} (2014) 037, [\href{http://arxiv.org/abs/1404.0008}{{\sf
  arXiv:1404.0008}}],
  [\href{http://dx.doi.org/10.1088/1475-7516/2014/06/037}{{\sf
  doi:10.1088/1475-7516/2014/06/037}}].

\bibitem{Yilmaz:2014lfa}
N.~T. Yilmaz, {\it {Effective Matter Cosmologies of Massive Gravity I:
  Non-Physical Fluids}},  {\sl JCAP} {\bf 1408} (2014) 037,
  [\href{http://arxiv.org/abs/1405.6402}{{\sf arXiv:1405.6402}}],
  [\href{http://dx.doi.org/10.1088/1475-7516/2014/08/037}{{\sf
  doi:10.1088/1475-7516/2014/08/037}}].

\bibitem{Konnig:2014xva}
F.~Koennig, Y.~Akrami, L.~Amendola, M.~Motta, and A.~R. Solomon, {\it {Stable
  and unstable cosmological models in bimetric massive gravity}},  {\sl
  Phys.Rev.} {\bf D90} (2014) 124014,
  [\href{http://arxiv.org/abs/1407.4331}{{\sf arXiv:1407.4331}}],
  [\href{http://dx.doi.org/10.1103/PhysRevD.90.124014}{{\sf
  doi:10.1103/PhysRevD.90.124014}}].

\bibitem{Cusin:2014psa}
G.~Cusin, R.~Durrer, P.~Guarato, and M.~Motta, {\it {Gravitational waves in
  bigravity cosmology}},  \href{http://arxiv.org/abs/1412.5979}{{\sf
  arXiv:1412.5979}}.

\bibitem{Lagos:2014lca}
M.~Lagos and P.~G. Ferreira, {\it {Cosmological perturbations in massive
  bigravity}},  {\sl JCAP} {\bf 1412} (2014) 026,
  [\href{http://arxiv.org/abs/1410.0207}{{\sf arXiv:1410.0207}}],
  [\href{http://dx.doi.org/10.1088/1475-7516/2014/12/026}{{\sf
  doi:10.1088/1475-7516/2014/12/026}}].

\bibitem{deRham:2014gla}
C.~de~Rham, M.~Fasiello, and A.~J. Tolley, {\it {Stable FLRW solutions in
  Generalized Massive Gravity}},  {\sl Int.J.Mod.Phys.} {\bf D23} (2014),
  no.~13 1443006, [\href{http://arxiv.org/abs/1410.0960}{{\sf
  arXiv:1410.0960}}], [\href{http://dx.doi.org/10.1142/S0218271814430068}{{\sf
  doi:10.1142/S0218271814430068}}].

\bibitem{Nersisyan:2015oha}
H.~Nersisyan, Y.~Akrami, and L.~Amendola, {\it {Consistent metric combinations
  in cosmology of massive bigravity}},
  \href{http://arxiv.org/abs/1502.03988}{{\sf arXiv:1502.03988}}.

\bibitem{Amendola:2015tua}
L.~Amendola, F.~Koennig, M.~Martinelli, V.~Pettorino, and M.~Zumalacarregui,
  {\it {Surfing gravitational waves: can bigravity survive growing tensor
  modes?}},  \href{http://arxiv.org/abs/1503.02490}{{\sf arXiv:1503.02490}}.

\bibitem{Mazuet:2015pea}
C.~Mazuet and M.~S. Volkov, {\it {De Sitter vacua in ghost-free massive gravity
  theory}},  \href{http://arxiv.org/abs/1503.03042}{{\sf arXiv:1503.03042}}.

\bibitem{Johnson:2015tfa}
M.~Johnson and A.~Terrana, {\it {Tensor Modes in Bigravity: Primordial to
  Present}},  \href{http://arxiv.org/abs/1503.05560}{{\sf arXiv:1503.05560}}.

\bibitem{Cusin:2015pya}
G.~Cusin, R.~Durrer, P.~Guarato, and M.~Motta, {\it {Inflationary perturbations
  in bimetric gravity}},  \href{http://arxiv.org/abs/1505.01091}{{\sf
  arXiv:1505.01091}}.

\bibitem{Akrami:2015qga}
Y.~Akrami, S.~Hassan, F.~Könnig, A.~Schmidt-May, and A.~R. Solomon, {\it
  {Bimetric gravity is cosmologically viable}},
  \href{http://arxiv.org/abs/1503.07521}{{\sf arXiv:1503.07521}}.

\bibitem{Fasiello:2015csa}
M.~Fasiello and R.~H. Ribeiro, {\it {Mild bounds on bigravity from primordial
  gravitational waves}},  \href{http://arxiv.org/abs/1505.00404}{{\sf
  arXiv:1505.00404}}.

\bibitem{Huang:2012pe}
Q.-G. Huang, Y.-S. Piao, and S.-Y. Zhou, {\it {Mass-Varying Massive Gravity}},
  {\sl Phys.Rev.} {\bf D86} (2012) 124014,
  [\href{http://arxiv.org/abs/1206.5678}{{\sf arXiv:1206.5678}}],
  [\href{http://dx.doi.org/10.1103/PhysRevD.86.124014}{{\sf
  doi:10.1103/PhysRevD.86.124014}}].

\bibitem{Leon:2013qh}
G.~Leon, J.~Saavedra, and E.~N. Saridakis, {\it {Cosmological behavior in
  extended nonlinear massive gravity}},  {\sl Class.Quant.Grav.} {\bf 30}
  (2013) 135001, [\href{http://arxiv.org/abs/1301.7419}{{\sf
  arXiv:1301.7419}}],
  [\href{http://dx.doi.org/10.1088/0264-9381/30/13/135001}{{\sf
  doi:10.1088/0264-9381/30/13/135001}}].

\bibitem{Cai:2012ag}
Y.-F. Cai, C.~Gao, and E.~N. Saridakis, {\it {Bounce and cyclic cosmology in
  extended nonlinear massive gravity}},  {\sl JCAP} {\bf 1210} (2012) 048,
  [\href{http://arxiv.org/abs/1207.3786}{{\sf arXiv:1207.3786}}],
  [\href{http://dx.doi.org/10.1088/1475-7516/2012/10/048}{{\sf
  doi:10.1088/1475-7516/2012/10/048}}].

\bibitem{Hinterbichler:2013dv}
K.~Hinterbichler, J.~Stokes, and M.~Trodden, {\it {Cosmologies of extended
  massive gravity}},  {\sl Phys.Lett.} {\bf B725} (2013) 1--5,
  [\href{http://arxiv.org/abs/1301.4993}{{\sf arXiv:1301.4993}}],
  [\href{http://dx.doi.org/10.1016/j.physletb.2013.07.009}{{\sf
  doi:10.1016/j.physletb.2013.07.009}}].

\bibitem{Wu:2013ii}
D.-J. Wu, Y.-S. Piao, and Y.-F. Cai, {\it {Dynamical analysis of the cosmology
  of mass-varying massive gravity}},  {\sl Phys.Lett.} {\bf B721} (2013) 7--12,
  [\href{http://arxiv.org/abs/1301.4326}{{\sf arXiv:1301.4326}}],
  [\href{http://dx.doi.org/10.1016/j.physletb.2013.02.055}{{\sf
  doi:10.1016/j.physletb.2013.02.055}}].

\bibitem{Bamba:2013aca}
K.~Bamba, M.~W. Hossain, R.~Myrzakulov, S.~Nojiri, and M.~Sami, {\it
  {Cosmological investigations of (extended) nonlinear massive gravity schemes
  with nonminimal coupling}},  {\sl Phys.Rev.} {\bf D89} (2014), no.~8 083518,
  [\href{http://arxiv.org/abs/1309.6413}{{\sf arXiv:1309.6413}}],
  [\href{http://dx.doi.org/10.1103/PhysRevD.89.083518}{{\sf
  doi:10.1103/PhysRevD.89.083518}}].

\bibitem{Cai:2013lqa}
Y.-F. Cai, F.~Duplessis, and E.~N. Saridakis, {\it {$F(R)$ nonlinear massive
  theories of gravity and their cosmological implications}},  {\sl Phys.Rev.}
  {\bf D90} (2014), no.~6 064051, [\href{http://arxiv.org/abs/1307.7150}{{\sf
  arXiv:1307.7150}}], [\href{http://dx.doi.org/10.1103/PhysRevD.90.064051}{{\sf
  doi:10.1103/PhysRevD.90.064051}}].

\bibitem{Cai:2014upa}
Y.-F. Cai and E.~N. Saridakis, {\it {Cosmology of F(R) nonlinear massive
  gravity}},  {\sl Phys.Rev.} {\bf D90} (2014), no.~6 063528,
  [\href{http://arxiv.org/abs/1401.4418}{{\sf arXiv:1401.4418}}],
  [\href{http://dx.doi.org/10.1103/PhysRevD.90.063528}{{\sf
  doi:10.1103/PhysRevD.90.063528}}].

\bibitem{D'Amico:2012zv}
G.~D'Amico, G.~Gabadadze, L.~Hui, and D.~Pirtskhalava, {\it {Quasidilaton:
  Theory and cosmology}},  {\sl Phys.Rev.} {\bf D87} (2013) 064037,
  [\href{http://arxiv.org/abs/1206.4253}{{\sf arXiv:1206.4253}}],
  [\href{http://dx.doi.org/10.1103/PhysRevD.87.064037}{{\sf
  doi:10.1103/PhysRevD.87.064037}}].

\bibitem{Gannouji:2013rwa}
R.~Gannouji, M.~W. Hossain, M.~Sami, and E.~N. Saridakis, {\it {Quasidilaton
  nonlinear massive gravity: Investigations of background cosmological
  dynamics}},  {\sl Phys.Rev.} {\bf D87} (2013) 123536,
  [\href{http://arxiv.org/abs/1304.5095}{{\sf arXiv:1304.5095}}],
  [\href{http://dx.doi.org/10.1103/PhysRevD.87.123536}{{\sf
  doi:10.1103/PhysRevD.87.123536}}].

\bibitem{Gumrukcuoglu:2013nza}
A.~E. G{\"u}mr{\"u}k\c{c}{\"u}o\u{g}lu, K.~Hinterbichler, C.~Lin, S.~Mukohyama,
  and M.~Trodden, {\it {Cosmological Perturbations in Extended Massive
  Gravity}},  {\sl Phys.Rev.} {\bf D88} (2013), no.~2 024023,
  [\href{http://arxiv.org/abs/1304.0449}{{\sf arXiv:1304.0449}}],
  [\href{http://dx.doi.org/10.1103/PhysRevD.88.024023}{{\sf
  doi:10.1103/PhysRevD.88.024023}}].

\bibitem{DeFelice:2013tsa}
A.~De~Felice and S.~Mukohyama, {\it {Towards consistent extension of
  quasidilaton massive gravity}},  {\sl Phys.Lett.} {\bf B728} (2014) 622--625,
  [\href{http://arxiv.org/abs/1306.5502}{{\sf arXiv:1306.5502}}],
  [\href{http://dx.doi.org/10.1016/j.physletb.2013.12.041}{{\sf
  doi:10.1016/j.physletb.2013.12.041}}].

\bibitem{Mukohyama:2014rca}
S.~Mukohyama, {\it {A new quasidilaton theory of massive gravity}},  {\sl JCAP}
  {\bf 1412} (2014), no.~12 011, [\href{http://arxiv.org/abs/1410.1996}{{\sf
  arXiv:1410.1996}}],
  [\href{http://dx.doi.org/10.1088/1475-7516/2014/12/011}{{\sf
  doi:10.1088/1475-7516/2014/12/011}}].

\bibitem{DeFelice:2013dua}
A.~De~Felice, A.~Emir~G{\"u}mr{\"u}k\c{c}{\"u}o\u{g}lu, and S.~Mukohyama, {\it
  {Generalized quasidilaton theory}},  {\sl Phys.Rev.} {\bf D88} (2013), no.~12
  124006, [\href{http://arxiv.org/abs/1309.3162}{{\sf arXiv:1309.3162}}],
  [\href{http://dx.doi.org/10.1103/PhysRevD.88.124006}{{\sf
  doi:10.1103/PhysRevD.88.124006}}].

\bibitem{Mukohyama:2013raa}
S.~Mukohyama, {\it {Extended quasidilaton massive gravity is ghost free}},
  \href{http://arxiv.org/abs/1309.2146}{{\sf arXiv:1309.2146}}.

\bibitem{Kahniashvili:2014wua}
T.~Kahniashvili, A.~Kar, G.~Lavrelashvili, N.~Agarwal, L.~Heisenberg, {\em
  et~al.}, {\it {Cosmic Expansion in Extended Quasidilaton Massive Gravity}},
  {\sl Phys.Rev.} {\bf D91} (2015), no.~4 041301,
  [\href{http://arxiv.org/abs/1412.4300}{{\sf arXiv:1412.4300}}],
  [\href{http://dx.doi.org/10.1103/PhysRevD.91.041301}{{\sf
  doi:10.1103/PhysRevD.91.041301}}].

\bibitem{Heisenberg:2015voa}
L.~Heisenberg, {\it {Revisiting perturbations in extended quasidilaton massive
  gravity}},  {\sl JCAP} {\bf 1504} (2015), no.~04 010,
  [\href{http://arxiv.org/abs/1501.07796}{{\sf arXiv:1501.07796}}],
  [\href{http://dx.doi.org/10.1088/1475-7516/2015/04/010}{{\sf
  doi:10.1088/1475-7516/2015/04/010}}].

\bibitem{Vegh:2013sk}
D.~Vegh, {\it {Holography without translational symmetry}},
  \href{http://arxiv.org/abs/1301.0537}{{\sf arXiv:1301.0537}}.

\bibitem{Blake:2013owa}
M.~Blake, D.~Tong, and D.~Vegh, {\it {Holographic Lattices Give the Graviton a
  Mass}},  {\sl Phys.Rev.Lett.} {\bf 112} (2014) 071602,
  [\href{http://arxiv.org/abs/1310.3832}{{\sf arXiv:1310.3832}}],
  [\href{http://dx.doi.org/10.1103/PhysRevLett.112.071602}{{\sf
  doi:10.1103/PhysRevLett.112.071602}}].

\bibitem{Blake:2013bqa}
M.~Blake and D.~Tong, {\it {Universal Resistivity from Holographic Massive
  Gravity}},  {\sl Phys.Rev.} {\bf D88} (2013), no.~10 106004,
  [\href{http://arxiv.org/abs/1308.4970}{{\sf arXiv:1308.4970}}],
  [\href{http://dx.doi.org/10.1103/PhysRevD.88.106004}{{\sf
  doi:10.1103/PhysRevD.88.106004}}].

\bibitem{Davison:2013jba}
R.~A. Davison, {\it {Momentum relaxation in holographic massive gravity}},
  {\sl Phys.Rev.} {\bf D88} (2013) 086003,
  [\href{http://arxiv.org/abs/1306.5792}{{\sf arXiv:1306.5792}}],
  [\href{http://dx.doi.org/10.1103/PhysRevD.88.086003}{{\sf
  doi:10.1103/PhysRevD.88.086003}}].

\bibitem{Zeng:2014uoa}
H.~B. Zeng and J.-P. Wu, {\it {Holographic superconductors from the massive
  gravity}},  {\sl Phys.Rev.} {\bf D90} (2014) 046001,
  [\href{http://arxiv.org/abs/1404.5321}{{\sf arXiv:1404.5321}}].

\bibitem{Siegel:1993sk}
W.~Siegel, {\it {Hidden gravity in open string field theory}},  {\sl Phys.Rev.}
  {\bf D49} (1994) 4144--4153, [\href{http://arxiv.org/abs/hep-th/9312117}{{\sf
  arXiv:hep-th/9312117}}],
  [\href{http://dx.doi.org/10.1103/PhysRevD.49.4144}{{\sf
  doi:10.1103/PhysRevD.49.4144}}].

\bibitem{ArkaniHamed:2002sp}
N.~Arkani-Hamed, H.~Georgi, and M.~D. Schwartz, {\it {Effective field theory
  for massive gravitons and gravity in theory space}},  {\sl Annals Phys.} {\bf
  305} (2003) 96--118, [\href{http://arxiv.org/abs/hep-th/0210184}{{\sf
  arXiv:hep-th/0210184}}],
  [\href{http://dx.doi.org/10.1016/S0003-4916(03)00068-X}{{\sf
  doi:10.1016/S0003-4916(03)00068-X}}].

\bibitem{Creminelli:2005qk}
P.~Creminelli, A.~Nicolis, M.~Papucci, and E.~Trincherini, {\it {Ghosts in
  massive gravity}},  {\sl JHEP} {\bf 0509} (2005) 003,
  [\href{http://arxiv.org/abs/hep-th/0505147}{{\sf arXiv:hep-th/0505147}}],
  [\href{http://dx.doi.org/10.1088/1126-6708/2005/09/003}{{\sf
  doi:10.1088/1126-6708/2005/09/003}}].

\bibitem{Deffayet:2005ys}
C.~Deffayet and J.-W. Rombouts, {\it {Ghosts, strong coupling and accidental
  symmetries in massive gravity}},  {\sl Phys.Rev.} {\bf D72} (2005) 044003,
  [\href{http://arxiv.org/abs/gr-qc/0505134}{{\sf arXiv:gr-qc/0505134}}],
  [\href{http://dx.doi.org/10.1103/PhysRevD.72.044003}{{\sf
  doi:10.1103/PhysRevD.72.044003}}].

\bibitem{Ostrogradski}
M.~Ostrogradksi, {\it {\emph{Memoires sur les equations differentielle
  relatives au probleme des isoperimetres.}}},  {\sl Mem. Ac. St. Petersbourg}
  {\bf \textbf{VI}} (1850) 385.

\bibitem{Nicolis:2008in}
A.~Nicolis, R.~Rattazzi, and E.~Trincherini, {\it {The Galileon as a local
  modification of gravity}},  {\sl Phys.Rev.} {\bf D79} (2009) 064036,
  [\href{http://arxiv.org/abs/0811.2197}{{\sf arXiv:0811.2197}}],
  [\href{http://dx.doi.org/10.1103/PhysRevD.79.064036}{{\sf
  doi:10.1103/PhysRevD.79.064036}}].

\bibitem{Vainshtein:1972sx}
A.~Vainshtein, {\it {To the problem of nonvanishing gravitation mass}},  {\sl
  Phys.Lett.} {\bf B39} (1972) 393--394,
  [\href{http://dx.doi.org/10.1016/0370-2693(72)90147-5}{{\sf
  doi:10.1016/0370-2693(72)90147-5}}].

\bibitem{vanDam:1970vg}
H.~van Dam and M.~Veltman, {\it {Massive and massless Yang-Mills and
  gravitational fields}},  {\sl Nucl.Phys.} {\bf B22} (1970) 397--411,
  [\href{http://dx.doi.org/10.1016/0550-3213(70)90416-5}{{\sf
  doi:10.1016/0550-3213(70)90416-5}}].

\bibitem{Zakharov:1970cc}
V.~Zakharov, {\it {Linearized gravitation theory and the graviton mass}},  {\sl
  JETP Lett.} {\bf 12} (1970) 312.

\bibitem{Babichev:2013usa}
E.~Babichev and C.~Deffayet, {\it {An introduction to the Vainshtein
  mechanism}},  {\sl Class.Quant.Grav.} {\bf 30} (2013) 184001,
  [\href{http://arxiv.org/abs/1304.7240}{{\sf arXiv:1304.7240}}],
  [\href{http://dx.doi.org/10.1088/0264-9381/30/18/184001}{{\sf
  doi:10.1088/0264-9381/30/18/184001}}].

\bibitem{Keltner:2015xda}
L.~Keltner and A.~J. Tolley, {\it {UV properties of Galileons: Spectral
  Densities}},  \href{http://arxiv.org/abs/1502.05706}{{\sf arXiv:1502.05706}}.

\bibitem{Ondo:2013wka}
N.~A. Ondo and A.~J. Tolley, {\it {Complete Decoupling Limit of Ghost-free
  Massive Gravity}},  {\sl JHEP} {\bf 1311} (2013) 059,
  [\href{http://arxiv.org/abs/1307.4769}{{\sf arXiv:1307.4769}}],
  [\href{http://dx.doi.org/10.1007/JHEP11(2013)059}{{\sf
  doi:10.1007/JHEP11(2013)059}}].

\bibitem{Gabadadze:2013ria}
G.~Gabadadze, K.~Hinterbichler, D.~Pirtskhalava, and Y.~Shang, {\it {Potential
  for general relativity and its geometry}},  {\sl Phys.Rev.} {\bf D88} (2013),
  no.~8 084003, [\href{http://arxiv.org/abs/1307.2245}{{\sf arXiv:1307.2245}}],
  [\href{http://dx.doi.org/10.1103/PhysRevD.88.084003}{{\sf
  doi:10.1103/PhysRevD.88.084003}}].

\bibitem{Noller:2015eda}
J.~Noller and J.~H.~C. Scargill, {\it {The decoupling limit of Multi-Gravity:
  Multi-Galileons, Dualities and More}},  {\sl JHEP} {\bf 1505} (2015) 034,
  [\href{http://arxiv.org/abs/1503.02700}{{\sf arXiv:1503.02700}}],
  [\href{http://dx.doi.org/10.1007/JHEP05(2015)034}{{\sf
  doi:10.1007/JHEP05(2015)034}}].

\bibitem{Hinterbichler:2013eza}
K.~Hinterbichler, {\it {Ghost-Free Derivative Interactions for a Massive
  Graviton}},  {\sl JHEP} {\bf 1310} (2013) 102,
  [\href{http://arxiv.org/abs/1305.7227}{{\sf arXiv:1305.7227}}],
  [\href{http://dx.doi.org/10.1007/JHEP10(2013)102}{{\sf
  doi:10.1007/JHEP10(2013)102}}].

\bibitem{Noller:2014ioa}
J.~Noller, {\it {On Consistent Kinetic and Derivative Interactions for
  Gravitons}},  \href{http://arxiv.org/abs/1409.7692}{{\sf arXiv:1409.7692}}.

\bibitem{Kimura:2013ika}
R.~Kimura and D.~Yamauchi, {\it {Derivative interactions in de
  Rham-Gabadadze-Tolley massive gravity}},  {\sl Phys.Rev.} {\bf D88} (2013)
  084025, [\href{http://arxiv.org/abs/1308.0523}{{\sf arXiv:1308.0523}}],
  [\href{http://dx.doi.org/10.1103/PhysRevD.88.084025}{{\sf
  doi:10.1103/PhysRevD.88.084025}}].

\bibitem{deRham:2015rxa}
C.~de~Rham, A.~Matas, and A.~J. Tolley, {\it {New Kinetic Terms for Massive
  Gravity and Multi-gravity: A No-Go in Vielbein Form}},
  \href{http://arxiv.org/abs/1505.00831}{{\sf arXiv:1505.00831}}.

\bibitem{Hassan:2012wr}
S.~Hassan, A.~Schmidt-May, and M.~von Strauss, {\it {On Consistent Theories of
  Massive Spin-2 Fields Coupled to Gravity}},  {\sl JHEP} {\bf 1305} (2013)
  086, [\href{http://arxiv.org/abs/1208.1515}{{\sf arXiv:1208.1515}}],
  [\href{http://dx.doi.org/10.1007/JHEP05(2013)086}{{\sf
  doi:10.1007/JHEP05(2013)086}}].

\bibitem{Yamashita:2014fga}
Y.~Yamashita, A.~De~Felice, and T.~Tanaka, {\it {Appearance of Boulware–Deser
  ghost in bigravity with doubly coupled matter}},  {\sl Int.J.Mod.Phys.} {\bf
  D23} (2014) 1443003, [\href{http://arxiv.org/abs/1408.0487}{{\sf
  arXiv:1408.0487}}], [\href{http://dx.doi.org/10.1142/S0218271814430032}{{\sf
  doi:10.1142/S0218271814430032}}].

\bibitem{deRham:2014naa}
C.~de~Rham, L.~Heisenberg, and R.~H. Ribeiro, {\it {On couplings to matter in
  massive (bi-)gravity}},  {\sl Class.Quant.Grav.} {\bf 32} (2015), no.~3
  035022, [\href{http://arxiv.org/abs/1408.1678}{{\sf arXiv:1408.1678}}],
  [\href{http://dx.doi.org/10.1088/0264-9381/32/3/035022}{{\sf
  doi:10.1088/0264-9381/32/3/035022}}].

\bibitem{Hassan:2014gta}
S.~Hassan, M.~Kocic, and A.~Schmidt-May, {\it {Absence of ghost in a new
  bimetric-matter coupling}},  \href{http://arxiv.org/abs/1409.1909}{{\sf
  arXiv:1409.1909}}.

\bibitem{deRham:2014fha}
C.~de~Rham, L.~Heisenberg, and R.~H. Ribeiro, {\it {Ghosts and matter couplings
  in massive gravity, bigravity and multigravity}},  {\sl Phys.Rev.} {\bf D90}
  (2014), no.~12 124042, [\href{http://arxiv.org/abs/1409.3834}{{\sf
  arXiv:1409.3834}}], [\href{http://dx.doi.org/10.1103/PhysRevD.90.124042}{{\sf
  doi:10.1103/PhysRevD.90.124042}}].

\bibitem{Noller:2014sta}
J.~Noller and S.~Melville, {\it {The coupling to matter in Massive, Bi- and
  Multi-Gravity}},  {\sl JCAP} {\bf 1501} (2015), no.~01 003,
  [\href{http://arxiv.org/abs/1408.5131}{{\sf arXiv:1408.5131}}],
  [\href{http://dx.doi.org/10.1088/1475-7516/2015/01/003}{{\sf
  doi:10.1088/1475-7516/2015/01/003}}].

\bibitem{Heisenberg:2014rka}
L.~Heisenberg, {\it {Quantum corrections in massive bigravity and new effective
  composite metrics}},  \href{http://arxiv.org/abs/1410.4239}{{\sf
  arXiv:1410.4239}}.

\bibitem{Hinterbichler:2015yaa}
K.~Hinterbichler and R.~A. Rosen, {\it {A Note on Ghost-Free Matter Couplings
  in Massive Gravity and Multi-Gravity}},
  \href{http://arxiv.org/abs/1503.06796}{{\sf arXiv:1503.06796}}.

\bibitem{deRham:2015cha}
C.~de~Rham and A.~J. Tolley, {\it {Vielbein to the Rescue?}},
  \href{http://arxiv.org/abs/1505.01450}{{\sf arXiv:1505.01450}}.

\bibitem{Akrami:2014lja}
Y.~Akrami, T.~S. Koivisto, and A.~R. Solomon, {\it {The nature of spacetime in
  bigravity: two metrics or none?}},  {\sl Gen.Rel.Grav.} {\bf 47} (2015) 1838,
  [\href{http://arxiv.org/abs/1404.0006}{{\sf arXiv:1404.0006}}],
  [\href{http://dx.doi.org/10.1007/s10714-014-1838-4}{{\sf
  doi:10.1007/s10714-014-1838-4}}].

\bibitem{Soloviev:2014eea}
V.~O. Soloviev, {\it {Bigravity in tetrad Hamiltonian formalism and matter
  couplings}},  \href{http://arxiv.org/abs/1410.0048}{{\sf arXiv:1410.0048}}.

\bibitem{Comelli:2015pua}
D.~Comelli, M.~Crisostomi, K.~Koyama, L.~Pilo, and G.~Tasinato, {\it {Cosmology
  of bigravity with doubly coupled matter}},
  \href{http://arxiv.org/abs/1501.00864}{{\sf arXiv:1501.00864}}.

\bibitem{Gumrukcuoglu:2015nua}
A.~E. G{\"u}mr{\"u}k\c{c}{\"u}o\u{g}lu, L.~Heisenberg, S.~Mukohyama, and
  N.~Tanahashi, {\it {Cosmology in bimetric theory with an effective composite
  coupling to matter}},  \href{http://arxiv.org/abs/1501.02790}{{\sf
  arXiv:1501.02790}}.

\bibitem{Gumrukcuoglu:2014xba}
A.~Emir~G{\"u}mr{\"u}k\c{c}{\"u}o\u{g}lu, L.~Heisenberg, and S.~Mukohyama, {\it
  {Cosmological perturbations in massive gravity with doubly coupled matter}},
  {\sl JCAP} {\bf 1502} (2015), no.~02 022,
  [\href{http://arxiv.org/abs/1409.7260}{{\sf arXiv:1409.7260}}],
  [\href{http://dx.doi.org/10.1088/1475-7516/2015/02/022}{{\sf
  doi:10.1088/1475-7516/2015/02/022}}].

\bibitem{Solomon:2014iwa}
A.~R. Solomon, J.~Enander, Y.~Akrami, T.~S. Koivisto, F.~Könnig, {\em et~al.},
  {\it {Cosmological viability of massive gravity with generalized matter
  coupling}},  {\sl JCAP} {\bf 1504} (2015), no.~04 027,
  [\href{http://arxiv.org/abs/1409.8300}{{\sf arXiv:1409.8300}}],
  [\href{http://dx.doi.org/10.1088/1475-7516/2015/04/027}{{\sf
  doi:10.1088/1475-7516/2015/04/027}}].

\bibitem{Enander:2014xga}
J.~Enander, A.~R. Solomon, Y.~Akrami, and E.~Mortsell, {\it {Cosmic expansion
  histories in massive bigravity with symmetric matter coupling}},  {\sl JCAP}
  {\bf 1501} (2015) 006, [\href{http://arxiv.org/abs/1409.2860}{{\sf
  arXiv:1409.2860}}],
  [\href{http://dx.doi.org/10.1088/1475-7516/2015/01/006}{{\sf
  doi:10.1088/1475-7516/2015/01/006}}].

\bibitem{Gao:2014xaa}
X.~Gao and D.~Yoshida, {\it {On coupling between Galileon and massive gravity
  with composite metrics}},  \href{http://arxiv.org/abs/1412.8471}{{\sf
  arXiv:1412.8471}}.

\bibitem{Deser:1983mm}
S.~Deser and R.~I. Nepomechie, {\it {Gauge Invariance Versus Masslessness In De
  Sitter Space}},  {\sl Ann. Phys.} {\bf 154} (1984) 396,
  [\href{http://dx.doi.org/10.1016/0003-4916(84)90156-8}{{\sf
  doi:10.1016/0003-4916(84)90156-8}}].

\bibitem{Deser:1983tm}
S.~Deser and R.~I. Nepomechie, {\it {Anomolous Propogation of Gauge Fields in
  Conformally Flat Spaces}},  {\sl Phys.Lett.} {\bf B132} (1983) 321,
  [\href{http://dx.doi.org/10.1016/0370-2693(83)90317-9}{{\sf
  doi:10.1016/0370-2693(83)90317-9}}].

\bibitem{Deser:2001pe}
S.~Deser and A.~Waldron, {\it {Gauge invariances and phases of massive higher
  spins in (A)dS}},  {\sl Phys.Rev.Lett.} {\bf 87} (2001) 031601,
  [\href{http://arxiv.org/abs/hep-th/0102166}{{\sf arXiv:hep-th/0102166}}],
  [\href{http://dx.doi.org/10.1103/PhysRevLett.87.031601}{{\sf
  doi:10.1103/PhysRevLett.87.031601}}].

\bibitem{Deser:2001us}
S.~Deser and A.~Waldron, {\it {Partial masslessness of higher spins in (A)dS}},
   {\sl Nucl.Phys.} {\bf B607} (2001) 577--604,
  [\href{http://arxiv.org/abs/hep-th/0103198}{{\sf arXiv:hep-th/0103198}}],
  [\href{http://dx.doi.org/10.1016/S0550-3213(01)00212-7}{{\sf
  doi:10.1016/S0550-3213(01)00212-7}}].

\bibitem{Deser:2001wx}
S.~Deser and A.~Waldron, {\it {Stability of massive cosmological gravitons}},
  {\sl Phys.Lett.} {\bf B508} (2001) 347--353,
  [\href{http://arxiv.org/abs/hep-th/0103255}{{\sf arXiv:hep-th/0103255}}],
  [\href{http://dx.doi.org/10.1016/S0370-2693(01)00523-8}{{\sf
  doi:10.1016/S0370-2693(01)00523-8}}].

\bibitem{Deser:2001xr}
S.~Deser and A.~Waldron, {\it {Null propagation of partially massless higher
  spins in (A)dS and cosmological constant speculations}},  {\sl Phys.Lett.}
  {\bf B513} (2001) 137--141, [\href{http://arxiv.org/abs/hep-th/0105181}{{\sf
  arXiv:hep-th/0105181}}],
  [\href{http://dx.doi.org/10.1016/S0370-2693(01)00756-0}{{\sf
  doi:10.1016/S0370-2693(01)00756-0}}].

\bibitem{Deser:2003gw}
S.~Deser and A.~Waldron, {\it {Arbitrary spin representations in de Sitter from
  dS / CFT with applications to dS supergravity}},  {\sl Nucl.Phys.} {\bf B662}
  (2003) 379--392, [\href{http://arxiv.org/abs/hep-th/0301068}{{\sf
  arXiv:hep-th/0301068}}],
  [\href{http://dx.doi.org/10.1016/S0550-3213(03)00348-1}{{\sf
  doi:10.1016/S0550-3213(03)00348-1}}].

\bibitem{Deser:2004ji}
S.~Deser and A.~Waldron, {\it {Conformal invariance of partially massless
  higher spins}},  {\sl Phys.Lett.} {\bf B603} (2004) 30,
  [\href{http://arxiv.org/abs/hep-th/0408155}{{\sf arXiv:hep-th/0408155}}],
  [\href{http://dx.doi.org/10.1016/j.physletb.2004.10.007}{{\sf
  doi:10.1016/j.physletb.2004.10.007}}].

\bibitem{Deser:2006zx}
S.~Deser and A.~Waldron, {\it {Partially Massless Spin 2 Electrodynamics}},
  {\sl Phys.Rev.} {\bf D74} (2006) 084036,
  [\href{http://arxiv.org/abs/hep-th/0609113}{{\sf arXiv:hep-th/0609113}}],
  [\href{http://dx.doi.org/10.1103/PhysRevD.74.084036}{{\sf
  doi:10.1103/PhysRevD.74.084036}}].

\bibitem{Deser:2013bs}
S.~Deser, E.~Joung, and A.~Waldron, {\it {Gravitational- and Self- Coupling of
  Partially Massless Spin 2}},  {\sl Phys.Rev.} {\bf D86} (2012) 104004,
  [\href{http://arxiv.org/abs/1301.4181}{{\sf arXiv:1301.4181}}].

\bibitem{Deser:2013uy}
S.~Deser, M.~Sandora, and A.~Waldron, {\it {Nonlinear Partially Massless from
  Massive Gravity?}},  \href{http://arxiv.org/abs/1301.5621}{{\sf
  arXiv:1301.5621}}.

\bibitem{deRham:2012kf}
C.~de~Rham and S.~Renaux-Petel, {\it {Massive Gravity on de Sitter and Unique
  Candidate for Partially Massless Gravity}},
  \href{http://arxiv.org/abs/1206.3482}{{\sf arXiv:1206.3482}}.

\bibitem{deRham:2013wv}
C.~de~Rham, K.~Hinterbichler, R.~A. Rosen, and A.~J. Tolley, {\it {Evidence for
  and Obstructions to Non-Linear Partially Massless Gravity}},  {\sl Phys.Rev.}
  {\bf D88} (2013) 024003, [\href{http://arxiv.org/abs/1302.0025}{{\sf
  arXiv:1302.0025}}], [\href{http://dx.doi.org/10.1103/PhysRevD.88.024003}{{\sf
  doi:10.1103/PhysRevD.88.024003}}].

\bibitem{Hassan:2012gz}
S.~Hassan, A.~Schmidt-May, and M.~von Strauss, {\it {On Partially Massless
  Bimetric Gravity}},  {\sl Physics Letters} {\bf B726,} (2013) 834,
  [\href{http://arxiv.org/abs/1208.1797}{{\sf arXiv:1208.1797}}].

\bibitem{Blanchet:2015sra}
L.~Blanchet and L.~Heisenberg, {\it {Dark Matter via Massive (bi-)Gravity}},
  \href{http://arxiv.org/abs/1504.00870}{{\sf arXiv:1504.00870}}.

\bibitem{deRham:2014tga}
C.~de~Rham, A.~Matas, N.~Ondo, and A.~J. Tolley, {\it {Interactions of Charged
  Spin-2 Fields}},  \href{http://arxiv.org/abs/1410.5422}{{\sf
  arXiv:1410.5422}}.

\bibitem{Hartnoll:2008kx}
S.~A. Hartnoll, C.~P. Herzog, and G.~T. Horowitz, {\it {Holographic
  Superconductors}},  {\sl JHEP} {\bf 0812} (2008) 015,
  [\href{http://arxiv.org/abs/0810.1563}{{\sf arXiv:0810.1563}}],
  [\href{http://dx.doi.org/10.1088/1126-6708/2008/12/015}{{\sf
  doi:10.1088/1126-6708/2008/12/015}}].

\bibitem{deRham:2013tfa}
C.~de~Rham, A.~Matas, and A.~J. Tolley, {\it {New Kinetic Interactions for
  Massive Gravity?}},  {\sl Class.Quant.Grav.} {\bf 31} (2014) 165004,
  [\href{http://arxiv.org/abs/1311.6485}{{\sf arXiv:1311.6485}}],
  [\href{http://dx.doi.org/10.1088/0264-9381/31/16/165004}{{\sf
  doi:10.1088/0264-9381/31/16/165004}}].

\bibitem{Huang:2015yga}
Q.-G. Huang, R.~H. Ribeiro, Y.-H. Xing, K.-C. Zhang, and S.-Y. Zhou, {\it {On
  the uniqueness of the non-minimal matter coupling in massive gravity and
  bigravity}},  \href{http://arxiv.org/abs/1505.02616}{{\sf arXiv:1505.02616}}.

\bibitem{Heisenberg:2015iqa}
L.~Heisenberg, {\it {More on effective composite metrics}},
  \href{http://arxiv.org/abs/1505.02966}{{\sf arXiv:1505.02966}}.

\bibitem{deRham:2014wfa}
C.~de~Rham and R.~H. Ribeiro, {\it {Riding on irrelevant operators}},  {\sl
  JCAP} {\bf 1411} (2014), no.~11 016,
  [\href{http://arxiv.org/abs/1405.5213}{{\sf arXiv:1405.5213}}],
  [\href{http://dx.doi.org/10.1088/1475-7516/2014/11/016}{{\sf
  doi:10.1088/1475-7516/2014/11/016}}].

\bibitem{Arnowitt:1962hi}
R.~L. Arnowitt, S.~Deser, and C.~W. Misner, {\it {The Dynamics of general
  relativity}},  {\sl Gen.Rel.Grav.} {\bf 40} (2008) 1997--2027,
  [\href{http://arxiv.org/abs/gr-qc/0405109}{{\sf arXiv:gr-qc/0405109}}],
  [\href{http://dx.doi.org/10.1007/s10714-008-0661-1}{{\sf
  doi:10.1007/s10714-008-0661-1}}].

\bibitem{deRham:2013hsa}
C.~de~Rham, M.~Fasiello, and A.~J. Tolley, {\it {Galileon Duality}},  {\sl
  Phys.Lett.} {\bf B733} (2014) 46--51,
  [\href{http://arxiv.org/abs/1308.2702}{{\sf arXiv:1308.2702}}],
  [\href{http://dx.doi.org/10.1016/j.physletb.2014.03.061}{{\sf
  doi:10.1016/j.physletb.2014.03.061}}].

\bibitem{deRham:2014lqa}
C.~De~Rham, L.~Keltner, and A.~J. Tolley, {\it {Generalized galileon duality}},
   {\sl Phys.Rev.} {\bf D90} (2014), no.~2 024050,
  [\href{http://arxiv.org/abs/1403.3690}{{\sf arXiv:1403.3690}}],
  [\href{http://dx.doi.org/10.1103/PhysRevD.90.024050}{{\sf
  doi:10.1103/PhysRevD.90.024050}}].

\bibitem{deRham:2013awa}
C.~de~Rham, A.~Matas, and A.~J. Tolley, {\it {Deconstructing Dimensions and
  Massive Gravity}},  {\sl Class.Quant.Grav.} {\bf 31} (2014) 025004,
  [\href{http://arxiv.org/abs/1308.4136}{{\sf arXiv:1308.4136}}],
  [\href{http://dx.doi.org/10.1088/0264-9381/31/2/025004}{{\sf
  doi:10.1088/0264-9381/31/2/025004}}].

\bibitem{Schmidt-May:2014xla}
A.~Schmidt-May, {\it {Mass eigenstates in bimetric theory with matter
  coupling}},  {\sl JCAP} {\bf 1501} (2015) 039,
  [\href{http://arxiv.org/abs/1409.3146}{{\sf arXiv:1409.3146}}],
  [\href{http://dx.doi.org/10.1088/1475-7516/2015/01/039}{{\sf
  doi:10.1088/1475-7516/2015/01/039}}].

\bibitem{Folkerts:2011ev}
S.~Folkerts, A.~Pritzel, and N.~Wintergerst, {\it {On ghosts in theories of
  self-interacting massive spin-2 particles}},
  \href{http://arxiv.org/abs/1107.3157}{{\sf arXiv:1107.3157}}.

\bibitem{Feynman:1996kb}
R.~Feynman, F.~Morinigo, W.~Wagner, and B.~Hatfield, {\em {Feynman lectures on
  gravitation}}.
\newblock Addison-Wesley, 1996.

\bibitem{Banados:2013fda}
M.~Banados, C.~Deffayet, and M.~Pino, {\it {The Boulware-Deser mode in 3D
  first-order massive gravity}},  {\sl Phys.Rev.} {\bf D88} (2013), no.~12
  124016, [\href{http://arxiv.org/abs/1310.3249}{{\sf arXiv:1310.3249}}],
  [\href{http://dx.doi.org/10.1103/PhysRevD.88.124016}{{\sf
  doi:10.1103/PhysRevD.88.124016}}].

\bibitem{Lovelock:1971yv}
D.~Lovelock, {\it {The Einstein tensor and its generalizations}},  {\sl
  J.Math.Phys.} {\bf 12} (1971) 498--501,
  [\href{http://dx.doi.org/10.1063/1.1665613}{{\sf doi:10.1063/1.1665613}}].

\bibitem{Deffayet:2010zh}
C.~Deffayet, S.~Deser, and G.~Esposito-Farese, {\it {Arbitrary $p$-form
  Galileons}},  {\sl Phys.Rev.} {\bf D82} (2010) 061501,
  [\href{http://arxiv.org/abs/1007.5278}{{\sf arXiv:1007.5278}}],
  [\href{http://dx.doi.org/10.1103/PhysRevD.82.061501}{{\sf
  doi:10.1103/PhysRevD.82.061501}}].

\bibitem{Padilla:2010de}
A.~Padilla, P.~M. Saffin, and S.-Y. Zhou, {\it {Bi-galileon theory I:
  Motivation and formulation}},  {\sl JHEP} {\bf 1012} (2010) 031,
  [\href{http://arxiv.org/abs/1007.5424}{{\sf arXiv:1007.5424}}],
  [\href{http://dx.doi.org/10.1007/JHEP12(2010)031}{{\sf
  doi:10.1007/JHEP12(2010)031}}].

\bibitem{Deffayet:2011gz}
C.~Deffayet, X.~Gao, D.~Steer, and G.~Zahariade, {\it {From k-essence to
  generalised Galileons}},  {\sl Phys.Rev.} {\bf D84} (2011) 064039,
  [\href{http://arxiv.org/abs/1103.3260}{{\sf arXiv:1103.3260}}],
  [\href{http://dx.doi.org/10.1103/PhysRevD.84.064039}{{\sf
  doi:10.1103/PhysRevD.84.064039}}].

\end{thebibliography}\endgroup

\end{document}